\documentstyle[prd,eqsecnum,aps]{revtex}
 
\def\be{\begin{equation}}
\def\ee{\end{equation}}
\def\ba{\begin{eqnarray}}
\def\ea{\end{eqnarray}}
\newcommand{\bec}{\begin{center}}
\newcommand{\eec}{\end{center}}

\begin{document}
%\preprint{Cologne-Mexico, gr-qc/yymmnnn}
%\draft
\input epsf

\title{TOPICAL REVIEW:\\ General relativistic boson stars}
 
\author{Franz E Schunck${\dagger}$ and Eckehard W Mielke${\ddagger}$}
 
\address{\small ${\dagger}$ Institut f\"ur Theoretische Physik,
Universit\"at zu K\"oln, 50923 K\"oln, Germany\\
E-mail: fs@thp.uni-koeln.de}
\address{\small ${\ddagger}$ Departamento de F\'{\i}sica,
Universidad Aut\'onoma Metropolitana--Iztapalapa,
Apartado Postal 55-534,\\ C.P. 09340, M\'exico, D.F., MEXICO\\
E-mail: ekke@xanum.uam.mx}
 
\date{\today ;\ Received XX YY 2003}

\maketitle

\begin{abstract}
{\bf Abstract.}
There is accumulating evidence that (fundamental) scalar fields
may exist in Nature.
The gravitational collapse of such a boson cloud would lead to a
{\em boson star} (BS) as a new type of a compact object.
Similarly as for white dwarfs and neutron stars, there exists
a limiting mass, below which a BS is {\em stable} against
complete gravitational collapse to a black hole.
 
According to the form of the self-interaction of the basic
constituents and the spacetime symmetry, we can distinguish
mini-, axidilaton, soliton, charged,
oscillating and rotating BSs. Their compactness
prevents a Newtonian approximation, however,
modifications of general relativity,
as in the case of Jordan-Brans-Dicke theory as a
low energy limit of strings,
would provide them with {\em gravitational memory}.
 
In general, a BS is a compact, {\em completely regular} configuration
with structured layers due to the anisotropy of scalar matter,
an exponentially decreasing 'halo',
a critical mass inversely proportional to constituent mass,
an effective radius, and a large particle number.
Due to the Heisenberg principle, there exists a completely stable branch,
and as a coherent state, it allows for rotating solutions with
{\em quantised angular momentum}.
 
In this review, we concentrate on the fascinating possibilities of
detecting the various subtypes of (excited) BSs:
Possible signals include gravitational redshift and (micro-)lensing,
emission of gravitational waves, or, in the case of a giant BS, its dark
matter contribution to the rotation curves of galactic halos.
\end{abstract}


{\footnotesize
\tableofcontents
}

\section{Introduction}
 
In this review, we will assume
that {\em fundamental scalar fields} exist in Nature,
that --- in the early stages of the universe ---
they would have formed absolutely stable soliton-type configurations
kept together by their self-generated gravitational field. Theoretically,
such configurations are known as {\em boson stars} (BSs).
In some characteristics, they resemble
neutron stars; in other aspects, they are different and,
thereby, astronomers may have a chance to distinguish BS signals from
other
compact objects, like neutron stars or black holes (BHs).
BSs can be regarded as descendants of self-gravitating photonic
configurations
called {\em geons} (gravitational electromagnetic units) and proposed in
1955
by Wheeler \cite{Wh55}.
 
In this review, we shall mainly concentrate on the results
which are not included in the first reviews from 1992
\cite{Je92,LP92,LM92} as well as later in the Marcel Grossman meetings in
Jerusalem and Rome \cite{mielkeMG8,MS02}.
Since then, the number of publications investigating how BSs could
possibly
be detected have increased;
we intend to focus strongly on these more recent research results.
This could provide astronomers a handling on possible signals of BSs.
All of these theoretical results are only a humble beginning of
the understanding where and in which scenarios a BS could be detected.
We hope other researchers may find this review stimulating for
gaining new ideas or for refining  older research.
 
A second intention of this review is to distinguish clearly the
different theoretical models underlying the label BS which could lead to
different observational consequences.
 
The first distinction is that the matter part of a BS can be
described by either a complex or by a real scalar field. Then, there can
be different interactions:
(a) self-interactions described by scalar field potentials,
(b) minimal coupling to gauge fields, the scalar field can be carrier
of a charge (electric or hypercharge, e.g.).
Moreover, the BS scalar field can couple,
in standard general relativity (GR), minimally,
or, in scalar-tensor (ST) or Jordan-Brans-Dicke (JBD) theory,
non-minimally to gravity.
In the latter case, if the strength of gravitational force
is influenced by the JBD scalar field, there is an interaction of the
BS scalar field with the real JBD scalar field; this can lead to
{\em gravitational memory} effects in BSs.
JBD theory is closely related to low energy limits of superstring
theories \cite{CJ78} which imply
the primordial production of scalar fields. Then, relics like
the dilaton, the axion, combined as {\em axidilaton}, or
other moduli fields could remain in our present epoch as
candidates.
 
Already the first two papers on BSs provided the two main directions:
The history of these {\em hypothetical} stars starts in 1968
with the work of Kaup \cite{K68} using a complex massive scalar field
with gravitational interactions in a semi-classical manner.
The energy-momentum tensor is calculated classically providing the
source for gravitation; a very detailed investigation of the solution
classes
was done in \cite{FLP87a}. However, one year later,
Ruffini and Bonazzola \cite{RB69} used field quantisation of a real
scalar field and considered the ground state of $N$ particles.
The vacuum expectation value of the field operators yield the same
energy-momentum tensor and thus, not surprisingly, the same
field equations. The two different physical constituents, complex versus
quantised real scalar field BS, yield the same macroscopical results.
It should be noted, however, that the gravitational field $g_{\mu\nu}$ is
kept
{\em classical} due to the non-renormalisability of standard GR, or,
alternatively, treated it as principal
low energy part of some renormalisable superstring model.
 
More recently, a BS using a field quantised complex scalar field has been
constructed. BSs consisting of pure semi-classical real scalars
cannot exist because their static solutions in flat spacetime are
unstable due to Derrick's theorem \cite{De64},
solutions may arise only if the real scalar field possesses a
time-dependence leading to non-static {\em oscillating} BSs.
However, if a fermion star is present as well, a real scalar component
can be added and if it interacts with the fermions, then a combined
boson-fermion star is the result as in the first calculation by
T.D.~Lee and Pang \cite{LP87}; cf.~Section \ref{leepang}.
 
A challenge for the BS model is that, so far, no {\em fundamental}
scalar particle has been
detected with certainty in the laboratory. Several are proposed by
theory:
The Higgs particle $h$ is a necessary ingredient of the standard model.
However, the possible discovery of the Higgs boson of
mass $m_{\rm h}=114.5$ GeV/$c^2$ at the Large Electron Positron (LEP)
collider at CERN \cite{Ac00,Ba00} gives the BS strand of
investigation a fresh impetus.
 
For the BS, we need a scalar particle which does not decay;
or if it decays (like the Higgs, e.g.), one has to assume that,
in the gravitational binding, the
inverse process is efficient enough for an equilibrium,
as is the case for the $\beta$-decay inside a neutron star.
In the latter case, the direct physical predecessor of that kind of a BS
is not clear; but, of course, unstable Higgs particles could not
have formed a BS by themselves.
In order to explore that unknown particle regime, one can
play with the model parameters such as particle mass or interaction
constants.
Then, BSs of rather different sizes can occur:
it could be just a `gravitational atom'; it could be as massive
as the presumed BH in the central part
of a galaxy; or it could be an alternative explanation
for parts of the dark matter in the halo of galaxies.
 
In 1995, experiments \cite{AK02} proved the existence of the fifth
possible
state of matter, the {\em Bose-Einstein condensate} (BEC);
in 2001, the Nobel prize was awarded for its experimental realization in
traps.
BSs, if they exist, would be an astrophysical realization with a
self-generated gravitational confinement, cf.~\cite{JB01,BLV01}.
 
Let us sum up some of the properties of complex scalar field BSs
(following an earlier version of \cite{T02}) in Table I
as we shall discuss in Section III.

\begin{center}
\parbox{14cm}
{TABLE I. Overview of some complex scalar field BS properties.}
\end{center}
%----------------------table beginning-------------------------
$$\vbox{\offinterlineskip
\hrule
\halign{&\vrule#&\strut\quad\hfil#\quad\hfil\cr
height2pt&\omit&&\omit&&\omit&\cr
& {\bf Property}      && {\bf BS}    &\cr
height2pt&\omit&&\omit&&\omit&\cr
\noalign{\hrule}
height2pt&\omit&&\omit&&\omit&\cr
& Constituents  \hfill
  &&   Scalars  (Bose-Einstein-Condensation)    \hfill &\cr
height2pt&\omit&&\omit&&\omit&\cr
& Pressure support \hfill
  &&   Heisenberg's uncertainty relation    \hfill &\cr
height2pt&\omit&&\omit&&\omit&\cr
& Size   \hfill
  &&  Gigantic up to very compact (few Schwarzschild radii): Table IV, Fig.~\ref{fig2} \hfill &\cr
height2pt&\omit&&\omit&&\omit&\cr
& Surface  \hfill
  &&   Atmosphere    \hfill &\cr
height2pt&\omit&&\omit&&\omit&\cr
& Appearance  \hfill
  &&  Transparent (if only gravitationally interacting)  \hfill &\cr
height2pt&\omit&&\omit&&\omit&\cr
& Structure  \hfill
  &&  Einstein-Klein-Gordon equation      \hfill &\cr
height2pt&\omit&&\omit&&\omit&\cr
& Gravitational potential  \hfill
  &&  Newtonian weak up to highly relativistic \hfill &\cr
height2pt&\omit&&\omit&&\omit&\cr
& Last stable orbit \hfill
  &&   None    \hfill &\cr
height2pt&\omit&&\omit&&\omit&\cr
& Rotation \hfill
  &&   Differentially; discrete    \hfill &\cr
height2pt&\omit&&\omit&&\omit&\cr
& Avoiding a baryonic BH by \hfill
  &&   Jet, particle dynamics    \hfill &\cr
height2pt&\omit&&\omit&&\omit&\cr
& Avoiding  a scalar BH by \hfill
  &&   Stability, evolution    \hfill &\cr
height2pt&\omit&&\omit&&\omit&\cr
& Gravitational redshift \hfill
  &&   Comparable to neutron star values, but larger due to transparency
\hfill &\cr
height2pt&\omit&&\omit&&\omit&\cr
& Gravitational (micro-)lensing \hfill
  &&   Extreme deflection angles possible (Fig.~\ref{fig3}); MACHOS \hfill &\cr
height2pt&\omit&&\omit&&\omit&\cr
& Luminosity  \hfill
  &&   Larger than luminosity of a BH   \hfill &\cr
height2pt&\omit&&\omit&&\omit&\cr
& Star disruption (tidal radius) \hfill
  &&   Yes    \hfill &\cr
height2pt&\omit&&\omit&&\omit&\cr
& Distinctive observational  \hfill
  &&   Broadening of emission lines    \hfill &\cr
& \ \ \ signatures    \hfill
  &&   Gravitational waves    \hfill &\cr
&     \hfill
  &&   {\v C}erenkov radiation    \hfill &\cr
height2pt&\omit&&\omit&&\omit&\cr
height2pt&\omit&&\omit&&\omit&\cr}
\hrule}$$
%----------------------table end-------------------------

\subsection{Fundamental scalar fields in Nature?}
 
The physical nature
of the  spin-0-particles out of which the BS is presumed to
consist, is still an open issue. Until now, no fundamental elementary
scalar particle has been detected with certainty in
accelerator experiments, which could serve as the main constituent
of the BS. In the theory of Glashow, Weinberg, and Salam,
a Higgs boson doublet $(\Phi^+, \Phi^0)$ and its anti-doublet
$(\Phi^-, \bar \Phi^0)$ are necessary ingredients in order
to generate masses for the `heavy photons', i.e.~the $W^{\pm }$
and $Z^0$ gauge vector bosons \cite{Qu83}.
After symmetry breaking, only one real scalar particle, the Higgs
particle
$h:=(\Phi^0 + \bar \Phi^0)/\sqrt{2}$, remains free and occurs in the
state
of a constant scalar field background \cite{Pe97}.
As it is indicated by
the rather heavy top quark \cite{Ab95} of 176 GeV/$c^2$,
the mass of the Higgs particle is expected to be below 1000 GeV/$c^2$.
This is supported by the possible discovery of the Higgs boson of
mass $m_{\rm h}=114.5$ GeV/$c^2$ at the Large Electron
Positron (LEP) collider at CERN \cite{Ac00,Ba00}.
In order to stabilise such a light Higgs against quantum fluctuations, a
{\em supersymmetric} extension
of the standard model is desirable. However, then it will be accompanied
by
additional heavy Higgs fields $H^0$, $A^0$, and a charged doublet
$H^{\pm}$
in the mass range of 100 GeV/$c^2$ to 1 TeV/$c^2$.
For the unlike case of a Higgs mass $m_{\rm h}$
above 1.2 TeV/$c^2$, the self-interaction $U(\Phi)$ of the
Higgs field is so large that any  perturbative approach of the
standard model becomes unreliable. Tentatively, a conformal extension of
the standard model with gravity included
has been analysed, cf.~\cite{PR94,HMMN95}.
Future high-energy experiments at the LHC at {\sc Cern} should reveal
if Higgs particles really exist in Nature.
 
As free particles, the Higgs boson of Glashow-Weinberg-Salam theory
is unstable, e.g.~with respect to the decays $h \rightarrow W^+ + W^-$
and
$h \rightarrow Z^0+ Z^0$, if it is heavier than the
gauge bosons. In a compact
object like the BS, these decay channels are expected to be in partial
equilibrium with the inverse processes $Z^0+ Z^0 \rightarrow h$ or,
via virtual $W$ triangle graph,
$Z^0 \rightarrow h + \gamma$, for instance \cite{CL84}, by utilising
gravitational binding energy. This is presumably in full analogy
with the neutron star \cite{HWW94,ST83,We99} or quark star
\cite{KWWG95,GKW95},
where one finds an equilibrium of $\beta $- and inverse $\beta $-decay
or of a quark-gluon plasma (with similar decay and
inverse-decay processes) and thus stability of the
macroscopic star with respect to radioactive decay.
 
The known spin-0-particles in particle physics, although effectively
described by a Klein-Gordon (KG) equation,
are not elementary, they consist of quarks having spin 1/2.
Nevertheless, they possess some physical properties which can help us to
understand which kind of particles can occur in the different BS types.
 
The electrically charged pions $\pi^{\pm}$ with a mass of about
140 MeV/$c^2$ are a typical example for
complex scalar fields \cite{IZ80,MS93}. The electrically neutral
${\rm K}^0$, $\bar{\rm K}^0$ mesons are described by a complex KG
field as well; ${\rm K}^0$ and $\bar{\rm K}^0$ are distinguished by their
hypercharge $Y:=B+S =\pm 1$, where $B=0$ is the baryon number
and $S$ is the strangeness corresponding to a global $U(1)$ symmetry.
In both cases,  particles (e.g.~$\pi^{+}$, ${\rm K}^0$)
and anti-particles (e.g.~$\pi^{-}$, $\bar{\rm K}^0$) occur. Generally,
Noether's theorem leads to a conserved charge
\be
Q = e ( N^+ - N^- )  \label{Ncharge}
\ee
related to the number $N$ of particles and anti-particles, respectively,
where $e$ defines the absolute value of the charge.
Let us stress that if the KG equation for a complex scalar field
admits a {\em global} $U(1)$ symmetry,
as in the case for the ${\rm K}^0$, $\bar{\rm K}^0$ mesons,
``merely'' a conserved particle number $N$ arises and the charge is
the total strangeness $S$. Regarding the high degree of instability
of pions, $K$ mesons, and Higgs fields, the effect of
gravitational binding is essential to see if
these particles had enough time to form a BS.
 
In string theories, there may exist several fundamental scalar
particles having different global or local charges.
For example, for complex scalar particles with global $U(1)$ symmetry or
for real scalar particles,
the BS charge shall be called just {\em boson number} $N$.
For complex scalar particles with local $U(1)$ symmetry,
the BS has a charge given by (\ref{Ncharge}).
In a BS with temperature near zero, we will assume that
all constituents are particles, i.e.~no anti-particles present, or vice
versa.
 
In his model,
Kaup \cite{K68} used a complex scalar field with global $U(1)$ symmetry
so that no gauge bosons are present.
Ruffini and Bonazzola \cite{RB69}, however, took a real
scalar field for the general relativistic BS and a complex scalar field
with global $U(1)$ symmetry for the Newtonian BS description.
In the case that $U(1)$ symmetry is {\em local},
the complex scalar field couples to the gauge field
contribution, cf.~\cite{IZ80}, p.~31;
Jetzer and van der Bij \cite{JB89} called their hypothetical
astrophysical object {\em charged boson star} (CBS).
Within the Glashow-Weinberg-Salam theory before symmetry-breaking
of the group $SU(2)\times U(1)$, this $U(1)$ charge
describes the weak hypercharge, and not an electric one.
During the earliest stages of the universe, a complex scalar field with
different kinds of $U(1)$ charge (than electric or hyperweak) could have
been generated.
 
There is also the particle/field classification with respect to the
spatial reflection $P$.
Both a real and a complex scalar field can be either a {\em scalar}
or a {\em pseudo-scalar}.
 
Thus, particle characterisation will lead to
different consequences for the detection of each BS model.

%*******************************************

\subsection{Boson star as a self-gravitating Bose-Einstein condensate
            \label{BEC}}
 
Since Einstein and Bose it is well-known that scalar fields represent
identical particles which
can occupy the same ground state. Such a {\em Bose-Einstein condensate}
(BEC) has been experimentally realized in 1995 for cold atoms of even
number of electrons, protons, and neutrons, see Anglin and Ketterle
\cite{AK02} for a recent review. In the mean-field ansatz, the
interaction
of the atoms in a dilute gas is approximated by the effective potential
\be
U(|\Psi|)_{\rm eff}= \frac{\lambda}{4} |\Psi|^4 \, .  \label{Ueff}
\ee
This leads to a {\em nonlinear} Schr\"odinger equation for $\Psi$,
in this context known as Gross-Pitaevskii equation.
In a microscopic approach, one introduces bosonic creation and
annihilation operators $b^\dagger$ and $b$, respectively, satisfying
\be
 [b,b^\dagger]  =1
\ee
and finds that every number conserving normal ordered correlation
function
$\langle b_1 \cdots b_n \rangle$ splits into the sum of all possible
products of contractions $\langle b_i  b_j \rangle$ as in
Wick's theorem of quantum field theory (QFT). For $n=1$
one recovers the Gross-Pitaevskii equation, whereas the next order leads
to the Hartree-Fock-Bogoliubov equations, see \cite{KB01} for details.
Therefore, it is gratifying to note that BSs with
{\em repulsive} self-interaction $U(|\Phi|)$ considered
already in Refs.~\cite{MS81,CSW86} have their counterparts in the
effective potential (\ref{Ueff}) of BEC.
Thus, some authors \cite{JB01,BLV01} advocate
consideration of a cold BS as a {\em self-gravitating} BEC on an
astrophysical scale.
 
Recently, the so-called {\em vortices}, collective excitations of BECs
with angular momentum in the direction of the vortex axis $z$,
have been predicted \cite{WH00} and then experimentally
prepared \cite{MCW00}. Since, quantum-mechanically, the z-component
$J_z= a N \hbar$ of the total angular momentum
is necessarily quantised by the {\em azimuthal} quantum number $a$,
it is not possible to ``continuously" deform this state to the ground
state,
and by this circumstance, contributing to its (meta-) stability.
In 1996, axisymmetric solutions of the Einstein-KG equations have
been found by us \cite{SM96,MS96,SM98,SM99} which are rotating BSs
(differentially due to the frame-dragging in curved spacetime)
and exhibit, as a collective state, the same relation $J =aN$ for the
total angular momentum as in the case of the vortices of an BEC and a
meta-stability against the decay into the ground state as well;
cf.~Section \ref{rotaBS}.

\section{Complex scalar field boson stars}
 
In this Section, in a nutshell, BSs are localised solutions of the
coupled system of
Einstein and general relativistic Klein-Gordon (KG) equations of a
complex scalar $\Phi$.
Depending on the self-interaction potential $U(|\Phi |^2)$,
BSs have received different labels: In 1968, Kaup
\cite{K68} referred to the localised solution of the linear KG equation
as a {\em Klein-Gordon geon}, whereas
in 1987, the same compact object was called {\em mini-soliton star}
\cite{FLP87a} within a series of publications
\cite{L87,L87b,FLP87a,FLP87b}
where also the terms {\em scalar soliton star}
or just {\em soliton star} were introduced which nowadays are coined
{\em non-topological soliton star}.
Before, in 1986, Colpi et al.~\cite{CSW86} baptised a self-gravitating
scalar field with $|\Phi|^4$ self-interaction a {\em boson star}.
So, in 1989, in a paper on stability, the mini-soliton star
was renamed to {\em mini-boson star} \cite{LP89}.
An additional $U(1)$ charge led to the label {\em charged BS}
\cite{JB89}.
In \cite{CLT00}, different BS models were summarised under
the label {\em scalar stars}. Let us stress that already Ruffini and
Bonazzola \cite{RB69} envisioned a star by calling their investigation
{\em systems of self-gravitating particles in GR}.
 
In his perspective paper, Kaup \cite{K68} had studied for the first time
the full general relativistic coupling of a linear complex KG
field to gravity for a localised configuration.
It is already realized that neither a Schwarzschild type
event horizon nor an initial singularity occurs in the corresponding
numerical solutions.
Moreover, the problem of the stability of
the resulting BS with respect to radial perturbations is treated.
It is shown that such objects are resistant to radial gravitational
collapse for a total mass $M$ below some {\em critical} mass and low
central density (related works include Refs.~\cite{D63,FM68,TWS75}).
The resulting configuration
is a macroscopic coherent state for which the KG field
can be treated as a semi-classical field.
 
The Lagrangian density of gravitationally coupled complex scalar
field $\Phi$ reads
\be
{\cal L}_{\rm BS} = \frac {\sqrt{\mid g\mid }}{2\kappa } \left \{ R
 + \kappa
   \Bigl [ g^{\mu \nu} (\partial_\mu \Phi^*) (\partial_\nu \Phi )
             - U(|\Phi |^2) \right ] \Bigr \} \, .
 \label{lagrBS}
\ee
Here $\kappa = 8\pi G$ is the gravitational constant in natural units,
$g$ the determinant of the metric $g_{\mu \nu }$,
$R:= g^{\mu \nu } R_{\mu \nu }
= g^{\mu \nu } \Bigl (
\partial_\nu \Gamma_{\mu \sigma }{}^\sigma
- \partial_\sigma \Gamma_{\mu \nu }{}^\sigma +
\Gamma_{\mu \sigma }{}^\alpha \Gamma_{\alpha \nu }{}^\sigma
- \Gamma_{\mu \nu }{}^\alpha \Gamma_{\alpha \sigma }{}^\sigma \Bigr )$
the curvature scalar with Tolman's sign convention \cite{T34}.
Using the principle of variation, one finds the coupled
Einstein-Klein-Gordon equations
\ba
 G_{\mu \nu }:=  R_{\mu \nu } - \frac{1}{2} g_{\mu \nu } R
             & = & -\kappa T_{\mu \nu } (\Phi ) \; , \label{phi152} \\
      \left (\Box + \frac {dU}{d\mid \Phi \mid^2} \right ) \Phi
             & = & 0 \; , \label{phi153} \\
      \left (\Box + \frac {dU}{d\mid \Phi \mid^2} \right ) \Phi^\ast
             & = & 0 \; , \label{phi154}
\ea
where
\be
T_{\mu \nu }(\Phi )
 = {1\over2} \Bigl [ (\partial_\mu \Phi^\ast )(\partial_\nu \Phi )
  + (\partial_\mu \Phi )(\partial_\nu \Phi^\ast ) \Bigr ]
  - g_{\mu \nu } {\cal L} (\Phi )/\sqrt {\mid g\mid }
 \label{tmunu}
\ee
is the {\em stress-ener\-gy ten\-sor} and
$
\Box := \left (1/\sqrt {\mid g\mid }\right )\,
 \partial_\mu  \left (\sqrt{\mid g\mid } g^{\mu \nu }
 \partial_\nu \right )
$
the generally covariant d'Alembertian.
 
The {\em stationarity} ansatz
\be
     \Phi (r,t) = P(r) e^{-i\omega t}   \label{statio}
\ee
and its complex conjugate describe a spherically symmetric bound
state of scalar fields with positive or negative frequency
$\omega$, respectively.
It ensures that the BS spacetime remains static.
(The case of a real scalar field can readily be accommodated in
this formalism, but requires $\omega=0$ due to $\Phi = \Phi^\ast$.)
Equations (\ref{phi153}) and (\ref{phi154}) give the same
differential equation.
BSs composed of complex scalars are symmetric under particle
anti-particle
conjugation $C$, i.e., there exists a degeneracy under the sign change
$\omega \rightarrow -\omega, N \rightarrow -N, e \rightarrow -e$,
cf.~Eq.~(\ref{teilchen2}), where $e$ denotes the $U(1)$ charge and $N:=
\int j^0 d^3x$ the particle number; cf.~Section \ref{critical}.
Hence, a BS in total consists of either particles or anti-particles
but the classical description in field equations above give us an
indication of its constituents by the sign of $N$. This persists also
in field quantisation where (\ref {Ncharge}) holds; cf.~Sections
\ref{Q-CSF-BS} and \ref{Q-RSF-BS}.
 
In the case of spherical symmetry, the general line-element can be
written
down as follows
\be
ds^2 = e^{\nu (r)} dt^2 - e^{\lambda (r)} \biggl [ dr^2 + r^2 \Bigl (
 d\theta^2 + \sin^2\theta d\varphi^2 \Bigr ) \biggr ] \; ,  \label{Schw}
\ee
in which the metric is static and the functions $\nu =\nu (r)$ and
$\lambda =\lambda (r)$ depend only on the Schwarzschild type radial
coordinate $r$.
{}From (\ref{phi152})-(\ref{phi153}),
a system of three coupled nonlinear equations for the
radial parts of the scalar and the (strong) gravitational tensor field
results
which has to be solved numerically.
 
Similarly as in the case of a prescribed Schwarzschild background
\cite{DM79}, the self-generated spacetime curvature affects from the
KG equation (\ref{phi153}) the resulting
{\em radial Schr\"o\-din\-ger equation}
\be
\left[\partial_{r*^2} -V_{\rm eff}(r^*) +\omega^2- m^2\right]P = 0
\ee
for the radial function $P(r):=\Phi e^{i\omega t}$
essentially via an {\em effective gravitational}
potential $V_{\rm eff}(r^*)= e^\nu dU/(d|\Phi|^2)
  + e^\nu l(l+1)/r^2
  + (\nu ' -\lambda ' )e^{\nu-\lambda}/2r $,
when written in terms of the tortoise coordinate
$r^*:=\int e^{(\lambda- \nu)/2}\, dr$, cf.~\cite{MS81}.
Then, it can be easily realized that localised solutions decrease
asymptotically as $P(r) \sim (1/r)\exp\left(-\sqrt{m^2 -\omega^2}\,
r\right)$
in a Schwarzschild-type asymptotic background.
As first shown by Kaup \cite{K68},
cf.~\cite{RB69} for the real scalar field case, metric and curvature
associated with a BS remain {\em completely regular}.

\subsection{Anisotropy of scalar matter}
 
That the stress-energy tensor (\ref{tmunu}) of a
BS, unlike a classical fluid, is in general {\em anisotropic}
was already noticed by Kaup \cite{K68}. For a spherically symmetric
configuration, it becomes diagonal,
i.e.~$T_\mu{} ^\nu(\Phi) = {\rm diag} \; (\rho , -p_r,$
$-p_\bot, -p_\bot )$ with
\ba
\rho &=& \frac{1}{2} (\omega^2 P^2 e^{-\nu} +P'^2 e^{-\lambda} +U )\; ,
\nonumber  \\
p_r &=&  \rho -  U \; , \;  \nonumber  \\
p_\bot &=&  p_r -  P'^2 e^{-\lambda } \; . \label{eqost1}
\ea
In contrast to a neutron star \cite{HT65,ZN71,ST83,St91},
where the ideal fluid approximation demands the
isotropy of the pressure, for spherically symmetric BSs
there are different stresses $p_r$ and $p_\bot$ in radial or
tangential directions, respectively.
Ruffini and Bonazzola \cite{RB69} display in their Fig.~3 the
anisotropy of the different stresses for a quantised real scalar
field BS. Since it
satisfies the same differential equations, we can use this figure
here as well. Gleiser \cite{Gl88} introduced
the notion of {\it fractional anisotropy}\ $a_f :=(p_r - p_\bot )/p_r =
P'^2 e^{-\lambda}/(\rho - U)$ which depends essentially
on the self-interaction. Furthermore,
the contracted Bianchi identity $\nabla^{\mu} T_{\mu}{}^{\nu}\equiv 0$
is equivalent to the equation
\be
 \frac{d}{dr} p_{\rm r}= -\nu^\prime\left(\rho + p_{\rm r} -\frac{2}{r}(
p_{\rm
 r}-p_\bot)\right)
\ee
of `hydrostatic' equilibrium for
an anisotropic fluid, a {\em generalisation of the
Tolman-Oppenheimer-Volkoff equation}, see Ref.~\cite{MS96}.
 
By defining the number density $n:=\omega e^{-\nu/2} P^2$ and the
differential pressure
$p_\Delta:=p_r-p_\bot =P'^2 e^{-\lambda }$,
Kaup found
\be
p_\Delta = p_\Delta(\rho,p_r,n) = \rho + p_r - n^2 m^2/(\rho-p_r)
  \label{eqost2}
\ee
as an {\em algebraic} equation of state, valid only for
the linear KG case. However,
the exact form of this algebraic relation depends on the
differential equations (\ref{phi152})-(\ref{phi153}). Consequently,
these four thermodynamical variables for the BS show a nonlocal behaviour
upon perturbations depending on the boundary conditions \cite{K68}.
Furthermore, Kaup derived that radial perturbations have to be
nonadiabatic.
 
{}For a spherically symmetric BS in its ground state
two different layers of the scalar matter are separated by
$p_\bot (R_{\rm c})=0$, i.e.~a zero of the tangential pressure.
Near the centre,
$p_\bot$ is positive and, after passing through zero at the
{\em core radius} $R_{\rm c}$, it
stays  negative until radial infinity \cite{Sch91}.
The core radius $R_{\rm c}$ is still {\em inside} the BS and
contains most of the scalar matter. Hence, all
three stresses are positive inside the BS core;
the boundary layer contains a matter distribution with
$p_r>0$ and $p_\bot<0$.

The choice of the potential $U$ implies a difference in the radial
pressure $p_r= \rho -U$, and, therefore, may
affect the physical properties of the BS.
In Table II, we shall indicate the main proposals of
current literature.

\begin{center}
\parbox{14cm}{TABLE II. Complex scalar field BS models distinguished by
the scalar self-interaction and year of the first publication.}
\end{center}
%----------------------table beginning-------------------------
$$\vbox{\offinterlineskip
\hrule
\halign{&\vrule#&\strut\quad\hfil#\quad\hfil\cr
height2pt&\omit&&\omit&&\omit&\cr
& {\bf Compact} && {\bf Self-Interaction} $U(|\Phi|^2 )$ && {\bf Year,
Publication} &\cr
&{\bf Object}  && [$\alpha,\beta,\lambda,\lambda_{(2n+2)},\Phi_0$
are constants] &&  &\cr
height2pt&\omit&&\omit&&\omit&\cr
\noalign{\hrule}
height2pt&\omit&&\omit&&\omit&\cr
& Mini-BS: \hfill && $U_{\rm K}=m^2|\Phi|^2$ \hfill
  && 1968, Kaup \cite{K68}  \hfill  &\cr
& Newtonian BS: \hfill && $U_{\rm N}=m^2 |\Phi|^2$ \hfill 
  && 1969, Ruffini-Bonazzola \cite{RB69} \hfill &\cr
& Self-interacting BS: \hfill
  &&$U_{\rm  HKG} = m^2|\Phi|^2 - \alpha |\Phi|^4 + \beta |\Phi|^6$
\hfill
  && 1981, Mielke-Scherzer \cite{MS81} \hfill &\cr
& BS: \hfill &&
  $U_{\rm CSW} = m^2|\Phi|^2 + \lambda |\Phi|^4/2 \hfill $
  && 1986, Colpi-Shapiro-Wasserman \cite{CSW86} \hfill  &\cr
& Non-topol.~Soliton Star: \hfill
  && $U_{\rm  NTS}=m^2 |\Phi|^2 (1 - |\Phi|^2/\Phi_0^2)^2 \hfill $
  && 1987, Friedberg-Lee-Pang \cite{FLP87b,LP92}  \hfill   &\cr
& General BS:  \hfill
  && $U_{\rm HKL} = U_{\rm CSW} + \ldots +
    \lambda_{(2n+2)} |\Phi|^{2n+2} \hfill $
  && 1999, Ho-Kim-Lee \cite{HKL99}  \hfill  &\cr
& Sine-Gordon BS:  \hfill
  && $U_{\rm SG} =
   \alpha  m^2 \bigl[\sin(\pi/2 [\beta \sqrt{|\Phi|^2}-1])+1\bigr] \hfill
$
  && 2000, Schunck-Torres \cite{ST00}  \hfill  &\cr
& Cosh-Gordon BS: \hfill &&
  $U_{\rm CG} =
  \alpha m^2 \bigl[\cosh(\beta \sqrt{|\Phi|^2}) - 1\bigr] \hfill $
  && 2000, Schunck-Torres \cite{ST00}  \hfill    &\cr
& Liouville BS: \hfill &&
  $U_{\rm L}=
  \alpha m^2 \bigl[\exp(\beta^2 |\Phi|^2)-1\bigr] \hfill $
  && 2000, Schunck-Torres \cite{ST00}  \hfill   &\cr
height2pt&\omit&&\omit&&\omit&\cr}
\hrule}$$
%----------------------table end-------------------------
The index of $U_{\rm HKG}$ in Table II stands for Heisenberg-Klein-Gordon.
As indicated, the scalar potential is a function of $|\Phi |^2$.
This means that the field $\Phi$ can be either a {\em scalar} or a
{\em pseudo-scalar}, depending whether for spatial reflections there
exist an unitary operator
${\cal P}$ for which ${\cal P}\Phi(x, t) {\cal P}^\dagger = \pm
\Phi(-x, t)$ holds.
For instance, the axion $a$ is a pseudo-scalar. The invariance of the
Lagrangian under ${\cal P}$ depends on whether or not the
self-interaction is even or odd.
It should be noticed that the
inclusion of $|\Phi|^6$ or higher order terms into the potential
implies that
the scalar part of the theory is no longer renormalisable.
Fig.~\ref{fig1} shows the Feynman diagram for a $|\Phi|^4$ interaction.
 
In the Mielke-Scherzer paper \cite{MS81}, actually the BS is
composed from several complex scalars which are in the same ground state
of a t'Hooft type {\em monopole configuration}
$\Phi^I \sim R(r)\, P^{|I|}_l(\cos\theta)$, where $I$ indicates
the $SO(2N)$ group index. In their calculation, an averaged
energy-momentum tensor is used as proposed by Power and Wheeler
\cite{PW57}, p.~488, deriving an angular momentum term in the field
equations.
 
Regarding the different BS labels, a historical remark is in order.
In 1984, Takasugi and Yoshimura \cite{TY84} calculated the gravitational
collapse of a {\em cold Bose gas} by using the Tolman-Oppenheimer-Volkoff
 equation and called their
result a {\em cold Bose star}; in the same year, the same name
was given to Ruffini and Bonazzola's BS \cite{BGZ84}.
The label {\em boson star} was first introduced by
Colpi et al.~in 1986 \cite{CSW86}.
The reason for that may have been the similarity of the mass
units with neutron star mass units which in both cases are
the order of magnitude
$M_{\rm Pl}^3/m^2$ (for details see Section \ref{critical}).
For if the scalar field mass $m$ is in the order of GeV/$c^2$,
we find for a BS the same mass range as that of a neutron star.
Due to this argument, Kaup's BS with mass unit $M_{\rm Pl}^2/m$
seems to be a small object, and it has been called mini-BS \cite{LP89}.
We would like to stress that the size and total mass of a BS depends
strongly on the unknown value of $m$ and on the issue whether there are
scalar self-interactions.
For corresponding $m$'s, even a so-called mini-BS may have
masses observed in active galactic nuclei;
cf.~Sections \ref{halo}, \ref{lensing}, \ref{macho}, \ref{galaxy}.
The existence of mini-BSs has also been proven mathematically
\cite{BW00}.

\subsection{Charged boson star \label{chargedBS}}
 
If the complex scalar field is {\em minimally} coupled to
local $U(1)$ gauge fields, the Lagrangian density necessarily
(cf.~\cite{IZ80}) adopts the form
\be
{\cal L}_{\rm CBS}={1\over 2\kappa }\sqrt{\mid g\mid }\; R
                  +{1\over 2} \sqrt{\mid g\mid}\;
   \Bigl [ g^{\mu \nu}(D_\mu \Phi )^\ast (D_\nu \Phi )
             - U(|\Phi|^2 ) \Bigr ]
   -{1 \over 4} \sqrt{\mid g\mid}\; F_{\mu \nu } F^{\mu \nu }
  \; . \label{lagrU1BS}
\ee
In accordance with the gauge principle, the coupling between the scalar
field and the $U(1)$-valued 1-form $A=A_\mu dx^\mu $, is introduced
via the gauge and, for scalar fields, general covariant derivative
$ D_\mu \Phi = \partial_\mu \Phi + i e A_\mu \Phi $,
where $e$ denotes the $U(1)$ coupling constant.
Furthermore, the Maxwell type term for the
two-form $F:= \frac{1}{2}F_{\mu \nu}
dx^\mu \wedge dx^\nu $ occurs, where
$F_{\mu \nu } = \partial_\mu A_\nu - \partial_\nu A_\mu $ is the
Faraday field strength.
 
\begin{figure}[h]
\centering
\leavevmode\epsfysize=2cm
\epsfbox{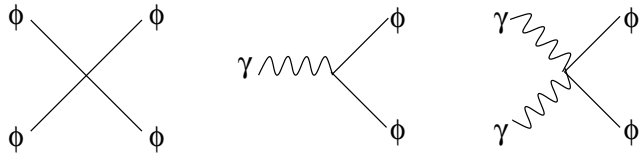}\\ \vskip 0.5cm
\caption[fig1]{\label{fig1} Feynman graph for the $\lambda |\Phi|^4$
self-interaction in Minkowski spacetime [left].
Feynman graphs for the
$e A^\mu \Phi^\ast \partial_\mu \Phi $
(or $e A^\mu \Phi \partial_\mu \Phi^\ast $, resp.) [middle]
and for the $e^2 A_\mu A^\mu |\Phi|^2$ [right]
self-interaction.}
\end{figure}

The expanded form
\be
(D_\mu \Phi )^\ast (D^\mu \Phi ) =
 (\partial_\mu \Phi )(\partial^\mu \Phi^\ast )
 + i e A^\mu \Phi \partial_\mu \Phi^\ast
 - i e A^\mu \Phi^\ast \partial_\mu \Phi
 + e^2 A_\mu A^\mu |\Phi|^2
\ee
of the kinetic term in (\ref{lagrU1BS}) helps us to understand
the gauge interactions in flat space-time for weak couplings.
In Fig.~\ref{fig1}, the two Feynman diagrams in order $e$ and order $e^2$
in the $U(1)$ gauge coupling are illustrated for flat spacetime.

Semi-classical solutions for (\ref{lagrU1BS}) have been found by
Jetzer and van der Bij \cite{JB89} for an electrically charged
complex scalar field. These spherically symmetric
configurations, called {\em charged BSs} (CBSs) \cite{JB89}, can only
exist
if the gravitational attraction is larger than the Coulomb repulsion,
i.e.~if $\alpha = e^2/4\pi < e^2_{\rm crit} = m^2/M^2_{\rm Pl}$,
where $\alpha$ is the fine structure constant;
cf.~the stability investigation in \cite{Je89b}. A macroscopic CBS
would accrete matter of opposite charge very rapidly or be
destroyed if the accreted matter consists of  anti-particles \cite{LM92}.
There may be vacuum instabilities induced by a CBS \cite{JLS93}
as we discuss now.
 
For particle-like CBSs, a screening mechanism can occur \cite{JLS93}.
If the CBS is more compact than the Compton
wavelength of the electron and supercritically charged, then
pair production of (i) electrons and positrons or of (ii) pions occurs.
For  fermionic pair-production, the critical CBS charge is about
$Z_c \sim 1/\alpha$, whereas for bosonic pairs, it is found that
$Z_c \sim 0.5/\alpha$. If a CBS has a supercritical charge, the
particles with opposing charge from the pair-production screen
every surplus above the critical charge
whereas their partners leave the newly formed
compound object. In \cite{JLS93} an example of a
CBS is given with 66 particles very concentrated at the centre, and a
cloud of 10 screening particles, e.g.~pions, almost outside of the
CBS ($m^2 =1000 m^2_\pi$); the screened CBS has a net charge of 56$e$.
There is also almost no influence on the physical properties of CBS
from the condensate (apart from its electromagnetic properties at large
distances). For CBSs with constituent mass $m$ comparable to $m_\pi$,
the vacuum is no longer overcritical, and no pair production occurs.
Furthermore, the pair production for the bosonic sector is three
order of magnitudes faster than for the fermionic one, so that
a pion condensate will most likely be created first. The CBS radius in
this example is less than 0.1 fermi. Furthermore,
it is a speculation that CHAMPs (CHArged Massive Particles) could
actually form CBSs, if such charged particles like pyrgons or maximons
would exist \cite{JLS93}.

A first-order perturbative approach
of a charged bosonic cloud has recently been examined \cite{DD02}
by which analytical formulae for mass, charge, and radius are achieved
if we have $\omega=m$, i.e., the cloud is in an energy state
even above the ones of excited CBS states.
This investigation is the starting point of the dynamics of
CBS formation as continued in Ref.~\cite{DD03};
cf.~Section \ref{waves}.

For the detection of CBSs, we should distinguish the different meanings
of a $U(1)$ symmetry. For example, within
the standard Glashow-Weinberg-Salam theory,
before symmetry-breaking of the group $SU(2)\times U(1)$, the $U(1)$
group describes the weak hypercharge. Hence, during
the earliest stages of the universe, a complex scalar field with
different kinds of $U(1)$ charge (than electric or hyperweak) could have
been generated. If we allow for a spontaneous symmetry breaking,
the Higgs field in the standard model becomes a real neutral
scalar field afterwards,
and possible mixing terms are eliminated in the unitary gauge.
Since, for a bound state like a BS, the weak coupling expansion does
not apply, in principle lattice field theory comes into force,
cf.~\cite{CL84}.

In a pre-CBS topic, stationary axisymmetric solutions are given in
\cite{BD77}.
In the physically different but mathematically similar context of the
Ginzburg-Landau system, cylindrically symmetric solutions for charged
scalar fields without self-gravitation
were found in \cite{OS96,OS97}.

%********************************************
\subsection{Critical masses of boson stars \label{critical}}
 
There is an essential distinction between real and complex scalar fields
in the {\em linear} KG equation:
\begin{itemize}
\item
The Lagrangian of a real scalar field is invariant only under the
{\em discrete} symmetry $\Phi \rightarrow \Phi'=-\Phi$,
\item
whereas the dynamics of a complex scalar  $\Phi:= (\Phi_1  +
i \Phi_2)/ \sqrt{2}$  is invariant under the global $O(2) \simeq U(1)$
symmetry $\Phi \rightarrow \Phi e^{i\alpha }$, where  $\alpha $ is a
constant.
\end{itemize}
 
The {\em Noether theorem} associates with each continuous
symmetry a {\em locally conserved current} $\partial_\mu j^\mu =0$ and
``charge'' (\ref{Ncharge}) which commutes with the Hamiltonian, i.e.
$[Q, H] =0$. For a complex scalar field,
the first ``constant of motion'' of our coupled system of equations
is given by the invariance of the Lagrangian density (\ref{lagrBS}) under
a global $U(1)$ transformation $\Phi \rightarrow \Phi e^{i\alpha }$ and
so, constraining the form of the self-interaction potential.
{}From the associated Noether current $j^\mu$, the
{\em particle number} $N$ arises:
\be
j^\mu = \frac {i}{2} \sqrt{\mid g\mid }\; g^{\mu \nu }
 [\Phi^\ast \partial_\nu \Phi -\Phi \partial_\nu \Phi^\ast ] \; ,
\qquad\qquad  N := \int j^0  d^3x  \; ;
\label{teilchen}
\ee
cf.~\cite{Go89,Go90}.
The second ``constant of motion'' arises from the Abelian group of time
translations, which gives rise to the concept of energy, cf.~\cite{Mi01}.
In a rest frame, this is proportional to
the {\em total gravitational mass} $M$ of localised solutions. Since
these
are asymptotically isolated configurations we can use
Tolman's expression \cite{T30}:
\be
M :=  \int (2T_0^{\; 0}-T_\mu^{\; \mu })
         \sqrt{\mid g\mid} \; d^3x \; . \label{Tolm}
\ee
In the case of a spherically symmetric CBS, where
the covector of the electromagnetic potential is given by
$A_\mu = [C(r),0,0,0]$, mass and particle number become
\ba
   M & = & 4\pi \int \limits_0^\infty \left (\rho +p_r +2p_\bot \right )
          e^{(\nu +\lambda )/2} r^2 dr  \\
     &  = & 4\pi \int \limits_0^ \infty
 \, \Bigl [ 2 P^2 (\omega +e C)^2 e^{-\nu} + C'^2 e^{-\lambda -\nu } - U
         \Bigr ] e^{(\nu +\lambda )/2} r^2 \; dr  \; , \\
N & = &  4 \pi \int\limits_0^\infty
    \; P^2 (\omega + e C ) e^{(\lambda -\nu )/2} r^2 \; dr  \; .
 \label{teilchen2}
\ea
Equivalently, the BS mass satisfies
$M=4\pi \int_0^\infty \rho \, r^2 dr$ which enters as a mass
parameter in the metric potentials
$e^{\nu(r)}, e^{-\lambda(r)} \rightarrow 1-2M/r$  of the vacuum
Schwarzschild solution, or in the asymptotic
regions where the BS matter tends to zero.
 
Since BSs are {\em macroscopic quantum states}, they are
prevented from complete
gravitational collapse by the Heisenberg uncertainty principle.
This provides us also with a crude mass estimate: For a boson to be
confined within the star of radius $R_0$,
the Compton wavelength of the scalar particle has to satisfy
$\lambda_\Phi= (2\pi\hbar/mc) \leq 2R_0$. On the other hand,
the star's radius should not be much less than the last stable Kepler
orbit
$3R_{\rm S}$ around a BH of Schwarzschild radius
$R_{\rm S}:= 2GM/c^2$ in order to avoid an instability
against complete gravitational collapse.
 
For the {\em mini}-BS, let us therefore assume an effective radius of
$R_0 \cong (\pi/2)^2 R_{\rm S} = 2.47 R_{\rm S}$. So far, the factor
$(\pi/2)^2$ has no theoretical foundation, but then the estimate
\be
M_{\rm crit} \cong (2/\pi)M_{\rm Pl}^2/m = 0.6366 \, M_{\rm Pl}^2/m
  \geq  M_{\rm Kaup} = 0.633\, M_{\rm Pl}^2/m\, ,
\label{Kauplim}
\ee
cf.~Ref.~\cite{Je92}, provides a rather good upper bound on the
{\em Kaup limit} $M_{\rm Kaup}$ which is the
numerically determined maximal mass of a {\em stable} BS.
Here $M_{\rm Pl}:=\sqrt{\hbar c /G}$ is the Planck mass
and $m$ the mass of a bosonic particle. As is typical for {\em solitonic}
solutions, the star becomes heavier for weaker coupling
$U(\vert\Phi\vert)$, i.e.~for smaller
constituent mass $m$ in a mini-BS.
 
In building star-sized BSs, one needs particles of ultra-low mass
$m$ or a non-linear \cite{MS81,CSW86,MS00} repulsive self-interaction
$U(\vert\Phi\vert)_{\rm CSW} = m^2\vert \Phi\vert^2\left(1 +
\tilde \Lambda \vert\Phi\vert^2/8\right)$, where
$\tilde \Lambda=4\lambda /m^2$.
%(not the same $\tilde \Lambda $ as in \cite{CSW86}).
In the latter case, for very small $\Lambda =\lambda M_{\rm Pl}^2/(4\pi
m^2)$,
the scalar field scales still as $\Phi \simeq M_{\rm Pl}/\sqrt{16\pi}$
and it turns out
\cite{MS00} that the Kaup limit (\ref{Kauplim})
can be applied again, but for a {\em rescaled} mass
$m\rightarrow m_{\rm resc}:=m/\sqrt{1 + \Lambda /8}$.
Then, in this regime, the maximal mass of a {\em stable} BS is
approximately
\be
M_{\rm crit} \simeq {2\over\pi}\sqrt{1 + \Lambda /8 \,}
\, {M_{\rm Pl}^2 \over m} \, .
\ee
For large $\Lambda$, the scalar field
scales as $\Phi \simeq M_{\rm Pl}/\sqrt{\Lambda}$, so that
$U(\vert\Phi\vert)_{\rm CSW} \simeq (1+2\pi) m^2 M_{\rm Pl}^2/\Lambda $;
the rescaled mass is now
$m\rightarrow m_{\rm resc} := m \sqrt{(1+2\pi)/\Lambda\, }$
and we find
\be
M_{\rm crit} \simeq {2\over\pi} \sqrt{\frac{\Lambda }{1+2\pi}}
{M_{\rm Pl}^2 \over m}
 = 0.236 \sqrt{\Lambda\,} {M_{\rm Pl}^2 \over m}
\, ;
\ee
cf.~Fig.~2 from Colpi et al.~\cite{CSW86} where, as well,
the value for the critical mass of 0.22
is derived in Eq.~(18). The complex scalar field
can be redefined as a dimensionless field
$\sigma=\sqrt{\kappa/2\,} \Phi=\sqrt{4\pi G\,} \Phi$
and the values of $\sigma$ for stable BSs are in the order of unity.
Then, depending on the constituent mass and the coupling constant,
the critical mass of a BS can reach the Chandrasekhar limit
$M_{\rm Ch}:=M_{\rm Pl}^3/m^2 \simeq M_\odot$ (if $m$ is about the
neutron mass), where $M_\odot$ denotes the mass of the sun,
and so can possibly even easily extend the limiting mass of 3.23
$M_\odot$ for rotating neutron stars \cite{MMS00}.

\begin{center}
\parbox{14cm}{TABLE III. Critical mass and particle number for
complex scalar field BS models.}
\end{center}
%----------------------table beginning-------------------------
$$\vbox{\offinterlineskip
\hrule
\halign{&\vrule#&\strut\quad\hfil#\quad\hfil\cr
height2pt&\omit&&\omit&&\omit&\cr
& {\bf Compact} && {\bf Critical Mass} && {\bf Particle Number } &\cr
&{\bf Object}  && $M_{\rm crit}$ && $N_{\rm crit}$ &\cr
height2pt&\omit&&\omit&&\omit&\cr
\noalign{\hrule}
height2pt&\omit&&\omit&&\omit&\cr
& Fermion Star: \hfill
  &&  $M_{\rm Ch}:=M_{\rm Pl}^3/m^2$ \hfill
  &&   $\sim(M_{\rm Pl}/m)^3$ \hfill &\cr
height2pt&\omit&&\omit&&\omit&\cr
& Mini-BS: \hfill
  &&  $M_{\rm Kaup}=0.633\, M_{\rm Pl}^2/m $ \hfill
  &&   $0.653 \,(M_{\rm Pl}/m)^2$ \hfill &\cr
height2pt&\omit&&\omit&&\omit&\cr
& BS with $U_{\rm CG}$ ($\beta=1$): \hfill
  &&  $M=0.638\, M_{\rm Pl}^2/m $ \hfill
  &&   $0.658 \,(M_{\rm Pl}/m)^2$ \hfill &\cr
height2pt&\omit&&\omit&&\omit&\cr
& BS with $U_{\rm SG}$ ($\beta=1$): \hfill
  &&  $M=0.620\, M_{\rm Pl}^2/m $ \hfill
  &&   $0.639 \,(M_{\rm Pl}/m)^2$ \hfill &\cr
height2pt&\omit&&\omit&&\omit&\cr
& Rotating mini-BS: ($a=1$) \hfill
  &&  $1.31\, M_{\rm Pl}^2/m $ \hfill
  &&  $1.38 \,(M_{\rm Pl}/m)^2$ \hfill &\cr
height2pt&\omit&&\omit&&\omit&\cr
& Rotating mini-BS: ($a=2$) \hfill
  &&  $\ge 2.21\, M_{\rm Pl}^2/m $ \hfill
  &&  $\ge 2.40 \,(M_{\rm Pl}/m)^2$ \hfill &\cr
height2pt&\omit&&\omit&&\omit&\cr
& BS: ($\lambda \gg 4\pi m^2/M^2_{\rm Pl}$) \hfill
  &&  $(1/\pi\sqrt{2\pi})\sqrt{\lambda} M_{\rm Pl}^3/ m^2$ \hfill
  &&  $\sim \sqrt{\lambda} (M_{\rm Pl}/m)^3$ \hfill &\cr
height2pt&\omit&&\omit&&\omit&\cr
& Non-topological Soliton Star: \hfill
  &&  $0.0198 (M_{\rm Pl}^4/ m \Phi_0^2)$ \hfill
  &&  $0.0055 (M_{\rm Pl}^5/ m^2 \Phi_0^3)$ \hfill &\cr
height2pt&\omit&&\omit&&\omit&\cr
& CBS: ($\lambda =0$) \hfill
  &&  $\sim 0.44/\sqrt{e_{\rm crit}-e}\; M_{\rm Pl}^2/ m$ \hfill
  &&  $\sim 0.44/\sqrt{e_{\rm crit}-e}\; (M_{\rm Pl}/ m)^2$ \hfill &\cr
& \hfill
  &&  $\sim M_{\rm Pl}^{3/2}/ m^{1/2}$ \hfill
  &&  $\sim (M_{\rm Pl}/ m)^{3/2}$ \hfill &\cr
height2pt&\omit&&\omit&&\omit&\cr
& CBS: ($\lambda \gg 4\pi m^2/M^2_{\rm Pl}$) \hfill
  &&  $\sim 0.226/\sqrt{e_{\rm crit}-e}\; \sqrt{\lambda}
          M_{\rm Pl}^3/ m^2$ \hfill
  &&  $\sim \sqrt{\lambda}/\sqrt{e_{\rm crit}-e}\;
         (M_{\rm Pl}/ m)^3$ \hfill &\cr
& \hfill
  &&  $\sim M_{\rm Pl}^{5/2}/ m^{3/2}$ \hfill
  &&  $\sim (M_{\rm Pl}/ m)^{5/2}$ \hfill &\cr
height2pt&\omit&&\omit&&\omit&\cr}
\hrule}$$
%----------------------table end-------------------------

In the case of a CBS (Table III),
the charges $e,e_{\rm crit}$ are measured in units
of order of magnitude $M_{\rm Pl}/m$ \cite{JB89},
so that CBSs have actually a {\em fractional mass law} \cite{Sch95}.
The non-topological soliton stars can easily have very large
masses; but compare the discussion of fermion soliton stars in
\cite{CV91a,CV91b} where the order of magnitude of a neutron star
can be found, cf.~Section \ref{leepang}.
 
In general (cf.~Table III), a BS is about the huge factor
$(M_{\rm Pl}/m)$
more massive than the mini-BS, as long as $(M_{\rm Pl}> m)$.
However, this depends also on the value
of the coupling constant $\lambda$, which even for the real Higgs scalar
of the standard model is experimentally quite unconstrained.
In astrophysical terms, the critical
mass of a stable BS is $M_{\rm crit} \cong 0.127 \sqrt{\lambda}\,
M_{\rm Ch}$ $=0.1 \sqrt{\lambda}$ ({\rm GeV/mc}$^2)^2\, M_\odot$
\cite{CSW86}.
For cosmologically relevant (invisible) axions (described by a real
pseudo-scalar $a$) of $m_{\rm a} \simeq 10^{-5}$ eV/$c^2$,
an axion star would have the ridiculously large mass of
$M_{\rm crit} \sim 10^{27}\sqrt{\lambda} M_\odot$ \cite{CSW86}.
 
In Table IV, we show several possible sizes which such localised
configurations could have.
We consider two models: on the left, the BS case with an
effectively large self-interaction of \cite{CSW86} where
$\lambda \gg 4\pi m^2/M^2_{\rm Pl}$; on the right hand side,
the self-interaction constant $\lambda$ does not fulfil this condition
or we have a mini-BS situation, for both cases, we can use the values of
$M$ and $R$ for $\lambda=1$.
The order of magnitude of the critical mass and radius
are given, hence, $R$ is calculated by the
Schwarzschild radius $2GM/c^2$ and so the BS configuration is effectively
general relativistic. Table IV provides examples of rather
different physical situations.
For example, in order to obtain a BS with about a solar mass
$M_\odot=10^{30}$ kg $=10^{57}$ GeV/$c^2$, a scalar field mass of
$m \sim 10^{-10}$ eV/$c^2$ for $\lambda = 0$, or $m \propto 1$ GeV/$c^2$
with
$\lambda = {\rm O}(1)$ is needed;
for the latter situation, we find the Colpi et al.~condition
$\lambda \gg 10^{-37}$, hence, the self-coupling can only be neglected
if it is extraordinarily tiny. In this example, the self-interacting
scalar particle has a mass comparable to a neutron, leading to a
BS with the dimensions of a neutron star, cf.~Section \ref{macho}.
Furthermore, we recognise that
the free scalar particle leads to a solar mass BS as well
so that it is not so ``mini" as its name mini-BS implies.
We give also values for a mean density $\bar \rho = M/R^3$.
If we reduce the self-interacting scalar mass further,
to $m \sim 1$ MeV/$c^2$, then we find a localised object that (for
$\lambda=1$ again)
has the size of the Sun but consists of $10^{6}$ solar masses;
this is reminiscent of supermassive black holes, for example as in active
galactic nuclei.
A mini-BS of the same dimensions would need an extremely
small scalar mass, which, for the time being, has no
observational support. By reducing $m$ further, we get a BS
which could play a role as a galactic halo, cf.~Section \ref{halo}.
 
Let us now increase $m$. With self-interaction and $m=100$ GeV/$c^2$
(about the possible
Higgs mass), the BS is just about 0.1 m but with a mass of $10^{26}$ kg;
for the mini-BS with the same bare mass, we find an ultra tiny atto-meter
BS, clearly below one Fermi. On the left hand side of this same
line, in order to derive the same
object with an effectively self-interaction constant, a very heavy
scalar particle is needed which is beyond experimental estimates
as well. Other not mentioned values of $m$ can easily be inter-
or extrapolated.

\begin{center}
\parbox{14cm}{TABLE IV.
Physical sizes of complex scalar field BS.}
\end{center}
%----------------------table beginning-------------------------
$$\vbox{\offinterlineskip
\hrule
\halign{&\vrule#&\strut\quad\hfil#\quad\hfil\cr
height2pt&\omit&&\omit&&\omit&\cr
& effective SI           && {\bf BS configuration}    && uneffective SI 
or
&\cr
& $U_{\rm CSW} \neq 0$   && $\leftarrow$ needs $\rightarrow$
  && $U_{\rm K} \neq 0$ &\cr
&       &&     && (take $\lambda=1$ in middle row) &\cr
height2pt&\omit&&\omit&&\omit&\cr
\noalign{\hrule}
height2pt&\omit&&\omit&&\omit&\cr
& $m=10^{10.5}$ GeV/$c^2$  \hfill
  && $M \approx 10^{9} \sqrt{\lambda}$ kg,
     $R \approx 1 \sqrt{\lambda}$ atto m, \hfill
      $\bar \rho \approx 10^{63}/\lambda$ kg/m$^3$
  &&   $m=100$ GeV/$c^2$      \hfill &\cr
height2pt&\omit&&\omit&&\omit&\cr
& $m=10^{6.5}$ GeV/$c^2$  \hfill
  && $M \approx 10^{17} \sqrt{\lambda}$ kg,
     $R \approx 1 \sqrt{\lambda}$ {\AA}, \hfill
      $\bar \rho \approx 10^{47}/\lambda$ kg/m$^3$
  &&   $m=1$ keV/$c^2$      \hfill &\cr
height2pt&\omit&&\omit&&\omit&\cr
& $m=10^{5}$ GeV/$c^2$  \hfill
  && $M \approx 10^{20} \sqrt{\lambda}$ kg,
     $R \approx 100 \sqrt{\lambda}$ nm, \hfill
      $\bar \rho \approx 10^{41}/\lambda$ kg/m$^3$
  &&   $m=1$ eV/$c^2$      \hfill &\cr
height2pt&\omit&&\omit&&\omit&\cr
& $m=100$ GeV/$c^2$  \hfill
  && $M \approx 10^{-4} \sqrt{\lambda} M_\odot$,
     $R \approx 0.1\phantom{^0} \sqrt{\lambda}$ m, \hfill
      $\bar \rho \approx 10^{29}/\lambda$ kg/m$^3$
  &&   $m=10^{-6}$ eV/$c^2$      \hfill &\cr
height2pt&\omit&&\omit&&\omit&\cr
& $m=1$ GeV/$c^2$  \hfill
  && $M \approx 10^{0\phantom{0}} \sqrt{\lambda} M_\odot$,
     $R \approx 10\phantom{^{00}} \sqrt{\lambda}$ km, \hfill
      $\bar \rho \approx 10^{18}/\lambda$ kg/m$^3$
  &&   $m=10^{-10}$ eV/$c^2$     \hfill &\cr
height2pt&\omit&&\omit&&\omit&\cr
& $m=1$ MeV/$c^2$  \hfill
  && $M \approx 10^{6\ } \sqrt{\lambda} M_\odot$,
     $R \approx 10^{6\ } \sqrt{\lambda}$ km, \hfill
      $\bar \rho \approx 10^{9\phantom{0}}/\lambda$ kg/m$^3$
  &&   $m=10^{-16}$ eV/$c^2$  \hfill &\cr
height2pt&\omit&&\omit&&\omit&\cr
& $m=1$ keV/$c^2$  \hfill
  && $M \approx 10^{12} \sqrt{\lambda} M_\odot$,
     $R \approx 10^{-1} \sqrt{\lambda}$ pc, \hfill
      $\bar \rho \approx 10^{0\phantom{0}}/\lambda$ kg/m$^3$
  &&   $m=10^{-22}$ eV/$c^2$    \hfill &\cr
height2pt&\omit&&\omit&&\omit&\cr
height2pt&\omit&&\omit&&\omit&\cr}
\hrule}$$
%----------------------table end-------------------------

A different approach to estimate the mass of a mini-BS is already
described in \cite{Je92}. There, analytic bounds are derived
on the ground state of a system of $N$ self-gravitating and
relativistic bosons in Newtonian approximation, without
distinguishing between real or complex scalars.
The Hamiltonian for an assembly of bosons is
\be
H = \sum_{i=1}^N \sqrt{{\bf p}_i^2+m^2}
   - \sum_{i>j} \frac{\kappa m^2}{|{\bf r}_{i}-{\bf r}_{j}|} \; ,
\ee
as implicitly used in  Chandrasekhar's calculation for
white dwarfs \cite{ST83}. The improved estimation in Ref.~\cite{RRS94}
for the lower and upper bounds on the maximal mini-BS mass is given by
\be
 0.8468 \, M_{\rm Pl}^2/m < M < 1.439 \, M_{\rm Pl}^2/m \; .
\ee
Since in the Hamiltonian general relativistic effects are ignored, the
bounds are above the Kaup limit $M_{\rm Kaup}$.

\subsection{Effective radius \label{radii}}
 
The surface and thus the radius
of a neutron star is characterised by a vanishing (radial) pressure
\cite{OV39,ST83}. In contradistinction to neutron stars, due to the
exponentially decreasing scalar field in a BS,
we can merely speak of an {\em exosphere} of a BS which corresponds to
the highest layer of a planet; alternatively, we can call it the
{\em halo} of a BS.
For an {\em effective radius} of a BS, several
proposals have been delivered in the literature.
 
The total mass $M$ can be used to define a radius where a certain
percentage
of mass is reached: $R=M(95\%)$ or even $R=M(99.9\%)$,
e.g.~\cite{DS00}. Alternatively, the energy-density $\rho$ is used in
\be
R_1:=\frac{\int_0^\infty \rho r^3 dr}{\int_0^\infty \rho r^2 dr}\,,
\ee
or with help of Tolman's expression (\ref{Tolm}) \cite{FLP87a}.
Taken the close relation to quantum mechanical
wave function into account, Mielke \cite{Mi79} proposed
the square root of the expectation value of the space operator as
BS radius
\be
R_2^2 := <|x|^2> = \frac{ \int_0^\infty |\Phi|^2 r^2 dr}{\int_0^\infty
|\Phi|^2 dr}, \quad
{\rm or} \quad R_3 :=
\frac{\int_0^\infty |\Phi|^2 r dr}{\int_0^\infty |\Phi|^2 dr}\,,
\ee
the latter being  even simpler.
All three definitions were investigated numerically in \cite{Sch91} for
spherically symmetric BSs with the result that
$R_1<R_2<R_3$ for one BS state; with increasing central densities
the BS radius decreases, there is a minimum at
$P(0)\sim 0.95 \sqrt{2/\kappa}$ and a maximum around
$P(0)\sim 1.5 \sqrt{2/\kappa}$. However, both these values are deep
in the unstable BS states.
 
Applying the conserved current (\ref{teilchen}), another definition
 \be
R_{\rm eff}:= \frac{1}{eN}\int_0^\infty j^0 r^3 dr
\ee
 was introduced in the case of CBSs, cf.  \cite{JB89}.
The reason is the long range nature of the electromagnetic force;
in this case, the definition $R_1$ would yield an infinite radius.
Let us notice that the Schwarzschild metric gives a general
lower bound on the radius of any type of object, namely $R > M/M_{\rm
Pl}^2$.
 
In all these definitions, the dimensions of the normalisation
and the integrated densities cancel each other,
except for the radial variable $r$. Therefore, the effective radius
of a mini-BS is proportional to  the Compton wavelength
$\lambda_\Phi= 2\pi\hbar/mc$ of the bosonic constituent.
This applies also to the case of a BS, except that we approximately
can use the rescaled mass, leading to
$\lambda_{\Phi_{\rm resc}} \simeq \sqrt{\Lambda }(2\pi\hbar/mc)$, which,
for large $\Lambda$, is about $(M_{\rm Pl}/ m)$ (a huge factor) times
larger. Thus, the corresponding effective radius $R_{\rm eff}$
justifies the distinction between a mini-BS and a BS.
 
The situation for a non-topological soliton star is different in this way
that it contains an interior part where the complex scalar field is
about the constant false vacuum value $\Phi_0$, a shell of width
$1/m$ over which the complex scalar field
decreases to zero, and the exterior vacuum.
The radius is then defined by the position of the shell.

\subsection{Stability and Heisenberg's uncertainty principle}

In the case of a ``normal'' star, the equation of state is approximately
that of an ideal gas, i.e.~the ratio of pressure $p$ to density $\rho$ is
a function of temperature. For a fermion star, as exemplified by
white dwarfs or neutron stars,
$p/\rho$ is independent of the temperature; for such degenerated matter,
the high pressure of the electron or neutron gas, respectively,
originates from Pauli's exclusion principle, resulting in a very
high Fermi energy.
 
Bosons are not susceptible to the exclusion principle; instead,
they can all occupy the same ground state. That bosons cannot
be localised within their Compton wavelength is guaranteed
by Heisenberg's uncertainty principle \cite{Gl88,Li76}.
 
In atoms as well as in a BS, the same principle applies and is the basic
warranty for its stability against collapse. From a macroscopic
point of view, our mathematical model shows that the macroscopic
scalar radial pressure works against gravity.
In a nutshell, a cold BS is a huge Bose-Einstein condensate, as we
demonstrated in more detail in Section \ref{BEC}.

\subsection{Stability, catastrophe theory, and critical
phenomena\label{catastrophe}}

According to Derrick's theorem \cite{De64,Ad02}, there are no
stable time-independent solutions of scalar fields in {\em flat} 4D
spacetime. One ingredient in order to circumvent this no-go theorem
is to allow for a {\em time-dependent} phase
factor $\exp(-i\omega t)$ in the stationarity ansatz (\ref{statio})
for complex scalars. For BSs, another factor is the attraction
of the gravitational field which
provides a counter-balance to the effective repulsion due to
kinetic energy and corresponding parts in the
self-interaction $U(\Phi)$ of the scalar field.
 
Moreover, for such soliton-type configurations kept together
by their self-generated gravitational field, the issue
of stability against {\em gravitational collapse} is crucial.
In the spherically
symmetric case, it was shown by Gleiser \cite{Gl88,GW89}, Jetzer
\cite{Je89,Je89b,Je89c,Je90}, as well as Lee and Pang \cite{LP89} that
BSs with small central densities and masses below the Kaup limit
are stable against small radial perturbations. (Kaup \cite{K68}
investigated
the stability with respect to radial perturbations as well but came
to a different answer.) Shortly afterwards, we have established
via {\em catastrophe theory} \cite{Ku89,KMS91,KMS91b,KS92,KS93,SKM92}
an even simpler way to prove {\em stability} of star solutions in
general.
The problem of non-radial pulsations of a BS was mathematically
formulated \cite{KYF91} and later quasinormal modes calculated
\cite{YEF94}; cf.~Section \ref{waves}.
 
Numerical simulations \cite{HC00,HC01} of a mini-BS together with a
surrounding spherical shell consisting of a massless real scalar
field have also led to qualitative stability predictions,
comparable to results in \cite{KMS91} as we show below.
In the simulation, the real scalar field implodes towards the mini-BS
centre; both fields are merely interacting gravitationally.
The pulse then passes through the origin, explodes and continues
to infinity. Meanwhile, the mini-BS is compressed into a state
which ultimately forms either a BH or disperses, depending on initial
data.
By varying the initial amplitude of the real scalar field,
a critical solution is defined as lying
between initial data resulting in BH formation and initial data
giving rise to dispersal; cf.~the numerical work on gravitational
collapse of a massless real scalar in \cite{GP87}.
 
As starting point of the numerical solution, a stable mini-BS solution
is taken, and perturbed by the
contribution of the real scalar field where two features are important:
(i) the initial amplitude, the so-called critical parameter $p$, and
(ii) the form of initial data (gaussian or kink). Because of
the choice of perturbation and a parameter value $p$ close to a critical
value $p^\star$, the stable mini-BS is transformed into an
intermediate oscillating unstable mini-BS state, the so-called
critical solution; cf.~\cite{Ch93}. The closer $p$ is
to $p^\star$, the more time the unstable mini-BS oscillates;
but, for $p<p^\star$, the mini-BS always ``explodes'' eventually,
i.e.~disperses all matter to infinity, leaving just a diffuse remnant
with low mass. For $p\ge p^\star$, the perturbed mini-BS forms
a BH afterwards, where a finite minimal BH mass is found. Due to this
and a scaling law relating the lifetime $\tau$ of
a near-critical solution to the proximity of the solution to the
critical point [$\tau \sim -\ln (p-p^\star)$], the critical solutions
belong to a so-called type I class.
 
The {\em critical mini-BS solutions} are a hybrid of an unstable
mini-BS solution and a small halo near the outer edge of the BS;
excited mini-BS states are not the explanation, they do differ
significantly.
The halo part seems to be just an artifact of the collision with the
real scalar field.
 
Additionally in \cite{HC00}, the numerical simulation method is used to
investigate and verify earlier results on the stability against
radial perturbations of mini-BS states with small and large central
densities.
The numerical simulation starts with a stable mini-BS solution.
If the critical mini-BS solution is reached, and if one ignores the
matter in its halo part, the rest is described by an unstable
mini-BS solution with less mass than the stable star had, but
larger central density value (see Fig.~16 in Ref. \cite{HC00}).
In fact, the numerical method of the linear perturbation theory
of \cite{GW89} is compared with the simulation method.
Both results coincide with each other, but, moreover, explicit values of
the frequencies of the fundamental mode and the first harmonic mode
are given. {}For every existing mode, there are either
two different stable mini-BS states,
or just one unstable one. Because the mode frequencies may have
observational consequences, we mention them in Section \ref{waves}.
Here the binding energy $M-mN$ has no
meaning \cite{HC00} in understanding the transitions in the
BS stability or the dynamics of the simulations.
 
These new simulations are related to the results
of an earlier paper \cite{KMS91} where the stability
was described qualitatively by catastrophe theory.
In numerical simulations, instability arises when the square of the mode
frequency is becoming
negative, i.e.~there exist an exponentially increasing mode,
and, by this, the mini-BS is destroyed. This starts to happen at extrema
in the diagrams ($M,P(0)$) or ($N,P(0))$,
cf.~Fig.~\ref{fig2} in this review;
these extrema belong to so-called cusps in the diagram ($M,N$),
also called {\em bifurcation diagram}.
By both the numerical investigation and the
catastrophe theory, one finds that, on the first cusp,
one stable mode becomes unstable; on the second cusp, a second
mode becomes unstable, and so on. In catastrophe theory,
the appearance of each cusp
is correlated to one Whitney surface (for one corresponding mode)
embedded in a higher dimensional parameter space.
The bifurcation diagram ($M,N$) arises from a particular
{\em projection} of this high dimensional parameter space.
In Fig.~2 of \cite{KMS91} which is the alternative bifurcation diagram
($M-mN,N$) using the binding energy instead of the BS mass,
three cusps are shown which are labeled $N_{C1}<N_{C2}<N_{C3}$;
$N_{C3}$ is the maximal particle number corresponding to the
onset of instability; $N_{C1}$ belongs to the second cusp
(or the next minimum following $N_{C3}$, resp.),
and $N_{C2}$ to the third cusp (or the next maximum following $N_{C1}$,
resp.). In Section VI.C of \cite{KMS91}, the physical
behaviour of non-equilibrium BSs is considered for $N_{C2}<N<N_{C3}$.
There, for constant BS mass, two BS states exist, one stable and one
unstable. Perturbed BSs will exhibit
oscillations around these two BS states before such a state
either collapses to a BH or disperses. This depends on
the strength of the perturbation, hence, on the parameter $p$
approaching its critical value from above.
 
In \cite{HC00}, numerical simulations for exactly this particle number
regime are carried out, and we see good agreements between the
simulations and the general predictions from catastrophe theory.
It would be interesting to see whether the numerical simulations
for the other regimes could be reconfirmed:
(a) for $N<N_{C1}$, only oscillating solutions (no collapse, no
dispersal), (b) for $N_{C1}<N<N_{C2}$, a pulsation behaviour of perturbed
BS states consisting of at least two different oscillation modes.

\subsection{Newtonian approximation - Gravitational atom I
\label{NewtBS}}

In a nut-shell, a BS is a stationary solution of a
(non-linear) KG equation in its own
gravitational field; cf.~\cite{Mi78,Mi80a}.
We treat this problem in a {\em semi-classical} manner,
because effects of the quantised gravitational field are neglected.
Therefore, a Newtonian BS was also called a {\it gravitational atom}
\cite{FG89}; hence, the excited states of a BS is the topic of this
subsection.
The results of the following subsection show that a
general relativistic BS
(i.e. one with strong self-gravity) as well can be designated as `atom'.
Thus, our `gravitational atoms' represent {\em coherent} quantum states,
which nevertheless can have macroscopic size and large masses.
The gravitational field is self-generated
via the energy-momentum tensor, but remains
completely classical, whereas the complex
scalar fields are treated to some extent as Schr\"odinger
wave functions, which in quantum field theory are
referred to as semi-classical as well.
 
Since a free KG equation for a complex scalar field is a
{\em relativistic generalisation} of the
{\em Schr\"odinger equation}, we consider for the ground state
a generalisation of the wave function \cite{H66,Mi81}
\ba
\vert N, n, l,a>: \qquad \Phi_{nla} (t,r,\theta,\varphi) &=& R_l^n(r)\,
 Y^{a}_l(\theta, \, \varphi)e^{-i (\omega_{nl}/\hbar)t} \nonumber\\
 &=& {1\over\sqrt{ 4\pi}} R_l^n(r)
 P^{a}_l(\cos\theta)\, e^{-i a \varphi}\,
e^{-i (\omega_{nl}/\hbar)t} \; ,  \label{Hatom}
\ea
where $R_l^n(r)$ gives the radial distribution,
$Y^{a}_l(\theta, \, \varphi)$ are the spherical harmonics,
$P^{a}_l(\cos\theta)$ are the
normalised Legendre polynomials, and $|a|\leq l$ are the quantum numbers
of {\em azimuthal} and {\em angular} momentum. Each mode has its
characteristic frequency $\omega_{nl}$.  As is well-known,
in the Newtonian limit Einstein's equation reduces to the
Poisson equation
\be
\Delta V = - \kappa \rho /2  \; ,  \label{Newt1}
\ee
where $V$ is Newton's gravitational potential,
$\rho = N m |\Phi|^2$ is the matter density, and $N$ the particle number.
Under the condition that all bosons are in the same state, the KG
equation reduces to
\be
\Delta \Phi_{nla} + 2m (\omega_{nl} + m V) \Phi_{nla} = 0 \; ,
\label{Newt2}
\ee
hence, a Schr\"odinger-type equation.
The relation for the conserved particle number
$
N = m \int |\Phi_{nla} |^2 r^2 \sin \theta dr d\theta d\varphi
$
gives us a normalisation condition on the wave function $R_l^n(r)$.
No self-interaction of the scalar field is introduced.
 
The coupled system of (\ref{Newt1}) and (\ref{Newt2}) has to be solved
numerically. A factorization as in (\ref{Hatom}) of the potential $V$ and
of
the scalar field $\Phi_{nla}$ yields two coupled ordinary
differential equations for the radial coordinate.
 
Several solutions have been derived so far.
The nodeless ground state with ($n=1,l=a=0$) was calculated first
by \cite{RB69} and can also be found in \cite{FG89} whereas in
\cite{FLP87a}
even Newtonian solutions up to 5 nodes are obtained. In Fig.~3 of
\cite{FG89}, we can see very nicely the difference in the Newtonian
and GR calculation. Additionally in \cite{FG89}, a BS was
calculated where scalar matter is distributed within the ground state
plus some matter in the lowest mode with nonzero quadrupole moment
(the $3d$ mode with $n=3$ and $l=2$) so that gravitational waves
can be radiated; cf.~Section \ref{waves}.
In \cite{SiSo95}, non-vanishing azimuthal angular momentum $a$ is
included,
and so, we have rotating Newtonian BSs; explicitly, solutions for
the combinations ($l=1, a=0$), ($l=1, a=1$), ($l=2, a=0$) are presented.
 
In cylindrical coordinates ($r,z$), the system (\ref{Newt1}),
(\ref{Newt2}) is an elliptic
boundary eigenvalue problem, and a finite element
method is used for solving the problem numerically in \cite{SB96}.
Due to the invariance of the differential equations with respect to
$z \rightarrow -z, V \rightarrow V, \Phi \rightarrow \pm \Phi $,
there are positive and negative parity classes where for the first
the spherically symmetric solution is derived.
Solutions with negative parity show two equatorial symmetric blobs
of scalar matter above and below the
equatorial plane, whereas the scalar field is equatorially
anti-symmetric.
This is comparable to a hydrogen atom in the
($n=2, l=1, a=0$) state, cf.~\cite{BD95}. Whereas this solution describes
an axisymmetric Newtonian BS, also non-axisymmetric Newtonian BSs
were obtained \cite{YE97c}. There, the scalar field is anti-symmetric
with respect to one of the following parity transformations:
(A) ($x,y,z$) $\rightarrow$ ($x,y,-z$), ($x,-y,z$);
(B) ($x,y,z$) $\rightarrow$ ($x,y,-z$), ($x,-y,z$), ($-x,y,z$).
The type A solutions show a BS with four blobs with maximal scalar
matter density in the ($x=0$)-plane. The type B solutions possess
eight blobs of scalar matter with zero density in the ($x=0$)-plane;
in both cases, the density vanishes also for $y=0$ and $z=0$.
 
In \cite{YE97}, the general relativistic extension of the
axisymmetric mini-BS solutions of
\cite{SB96} were exhibited. These negative parity solutions
are classified by the derivative $\Phi_r (r=0, \theta=0)$.
The critical mass is given by $M_{\rm crit}=1.05 M_{\rm Pl}^2/m$.
 
An analytical solution for a  BS in the limit
$\Lambda \gg 1$ is determined in \cite{LK96} for the rescaled scalar
field
$
\sigma_\ast (x_\ast) =
 \sqrt{(\gamma_0 \sin \left ( \sqrt{2\,}x_\ast \right )/\sqrt{2\,}x_\ast
\;}$,
where $\gamma_0$ is the absolute central value of the Newtonian
potential.

\subsection{Boson stars - Gravitational atom II}
 
In the case of the general relativistic BS, several excited
BS states have been computed.
The first solutions with nodes, i.e.~principal quantum number
$n>1$, were determined by Mielke and Scherzer \cite{MS81}
who, motivated by Heisenberg's non-linear spinor equation
\cite{H66,Mi81},
added self-inter\-acting terms describing the interaction
between the bosonic particles in the BS configuration.
 
The work of Friedberg et al.~\cite{FLP87a} investigated
solutions with up to 56 nodes within the mini-BS scalar field.
Solutions with nodes possess a similar behaviour in the
mass-particle-number
function as solutions without nodes, i.e.~a maximal mass or cusps in the
mass-particle-number diagram were found.
For the first cusp,  the maximal
mass depends linearly \cite{FLP87a} on the particle number.
There as well, the solutions with zero nodes are compared with
solutions with ten nodes.
 
Recently, {\em phase-shifted BS} solutions have been constructed
\cite{HC03}. Normally, the BS complex scalar field
$\Phi (r,t)= P(r) \cos (\omega t) + i P(r) \cos (\omega t + \delta)$
has a phase of $\delta=\pi/2$ between its real and its imaginary part.
For $\delta \neq \pm \pi/2$, such solutions are investigated and show
a BS like behaviour. Long-term numerical evolutions indicate
that phase-shifted BSs are stable but periodic; mass-energy is
exchanged between the real and imaginary fields. It seems that
BSs can be more generic than previously thought.
For $\delta =0$, the oscillating BSs are approximated \cite{SeSu91}.

\subsection{Rotating boson stars - Gravitational atom III \label{rotaBS}}
 
In Refs.~\cite{SM98,SM99b,Sch95,SM96,MS96}, we proved numerically that
regular
stationary and axisymmetric solutions, which can be regarded as
{\em rapidly rotating} BSs, exist in GR.
In the Einstein-KG system (\ref{phi152}), (\ref{phi153}),
we use an isotropic stationary {\em axisymmetric} line element
\be
ds^2 = f(r,\theta ) dt^2 - 2 k(r,\theta ) dt d\varphi
- l(r,\theta ) d\varphi ^2
- e^{\mu (r,\theta )} \left ( dr^2 + r^2 d\theta ^2 \right )
\; , \label{isokerr}
\ee
where $f,k,l,\mu$ are some functions to be determined numerically.
In order to find rotating BSs in ground state, the scalar field ansatz is
\be
\Phi (r,\theta ,\varphi ,t) :=
  P(r,\theta ) e^{-i a \varphi} e^{-i \omega t} \; ,
\label{stattt}
\ee
where the uniqueness of the scalar field under a complete rotation
$\Phi(\varphi) =\Phi(\varphi +2\pi)$ requires
$a=0, \pm 1, \pm 2, \ldots $, as in quantum mechanics.
For the resulting nonlinear elliptical system, we applied a finite
difference
method and found solutions for the mini-BS with $a$ up to 500.
Sequences of rotating mini-BSs with $a=1$ and $a=2$ were constructed in
\cite{YE97b} where every BS is characterised by the maximal value of the
scalar field which is usually not on the rotation axis. For $a=1$, the
energy
density has a non-vanishing value on the rotation axis,
whereas, for $a\ge2$, it has to vanish identically there,
i.e., vacuum predominates; cf.~the situation of the rotating BS
in (2+1)-dimensions in Section \ref{2+1}.
The maximum mass increases above the Kaup limit:
for $a=1$, it is $1.31 M_{\rm Pl}^2/m$
and, for $a=2$, it is at least $2.21 M_{\rm Pl}^2/m$.
In \cite{M99}, rotating BSs were calculated
with a numerically faster multigrid method.
 
The total angular momentum $J$ of a BS is given by
\be
J = a N \; , \label{dreh}
\ee
i.e.~a BS can change its rotational state only in discrete
steps resembling gravitational atoms. Hence,
the angular momentum $J$ is quantised by the azimuthal quantum number
$a$.
The lifting of the angular momentum degeneracy, familiar from
quantum mechanics, is a {\em gravito-magnetic} effect \cite{Ma74}
due to the rotating frame. It is,
however, surprising that this still holds for macroscopic
configurations. Eventually, it is now understandable why no slowly
rotating BS
states
near the spherically symmetric ones could be found \cite{KKF94}.
 
If we imagine what happens if a BS starts to rotate, we find out that
the BS rearranges its structure from the sphere to a torus.
The BS in this situation looks like the ($n=2,l=1,a=1$) state of
a hydrogen atom, cf.~\cite{BD95}.
Qualitatively, due to the finite velocity of light and the infinite
range of the scalar matter within the BS, it follows that the form of
rotation has to be differential, not uniform.
The rotating solutions of the coupled
Einstein-KG equations represent the field-theoretical pendant of
rotating neutron stars which have been studied numerically
for various equations of state, different approximation schemes
\cite{FrI92,CST94,Er93} as well as with quadrupole moments \cite{MMS00}
as a model for (millisecond) pulsars.
 
A general insight into the physical characteristics of rotating BS
matter can be obtained from the energy-momentum tensor \cite{Sch95}
\\

\be
T_\mu{}^\nu = \pmatrix{
  T_t{}^t       & 0            & 0                 & T_t{}^\varphi \cr
  0             & T_r{}^r      & T_r{}^\theta      & 0 \cr
  0             & T_\theta{}^r & T_\theta{}^\theta & 0 \cr
  T_\varphi{}^t & 0            & 0                 & T_\varphi{}^\varphi \cr
}   \; ,
\ee
in which the shear tensions $T_\theta{}^r$ and
$T_r{}^\theta$, as well as all main tensions are different from each
other,
i.e.~$T_r{}^r \neq T_\theta{}^\theta \neq T_\varphi{}^\varphi$, and
non-vanishing. Compared to a rotating ideal fluid of a neutron star
with $T^{\mu \nu } = (\rho + p) u^\mu u^\nu - p g^{\mu \nu }$,
there are no shear tensions at all
and the main stresses generally obey
$T_r{}^r = T_\theta{}^\theta \neq T_\varphi{}^\varphi$.
 
A rotating  BS, i.e.~with large self-interaction constant
$\lambda \gg 4\pi m^2/M^2_{\rm Pl}$, was investigated by Ryan \cite{R97}.
For the construction of the solutions,
the fact that the derivatives of the scalar field vanish (due to the
scaling with $\lambda$) yields an ideal fluid of BS matter as in
the non-rotating case \cite{CSW86}. From the solutions, multipole
moments like mass, angular momentum, and mass quadrupole moment
are determined. For fixed angular momentum $J$,
the maximal mass was calculated by increasing the maximal scalar field
value until the numerical computation became unstable (Fig.~2 in
\cite{R97});
under this condition, the maximal mass is about
0.58 $\sqrt{\Lambda} M^2_{\rm Pl}/m^2$ in our units which are
different by a factor $\sqrt{4\pi\,}$ to the ones in \cite{R97};
further details in Section \ref{waves}.
 
More recently, in flat spacetime soliton rings are constructed
\cite{AFKP01}
in a similar way from a complex scalar. They are {\em metastable}
due to the same combination of conserved Noether charge and angular
momentum.

%***************************************
 
\subsection{Formation}
 
It has been demonstrated that BSs can actually form from a
primordial bosonic ``cloud" by Jeans instability
\cite{KMZ85,FG89,Gr90,BGR90,ML90,T91,Je92,LM92,HBG00,RT00}.
(The primordial formation of non-gravitating non-topological soliton
solutions was studied by Frieman et al.~\cite{FGGK88,FOGA89}.)
That the quantum aspects of this ground state instability of the
complex scalar field acts like a classical Jeans instability is shown
in \cite{Gr90b}. The formation of a spherically symmetric mini-BS
as a Bose-Einstein condensation is considered in \cite{BiH00}.
One of the first steps in the formation process
of CBSs is investigated in \cite{DD02,DD03}
by calculating the dynamics of charged bosonic clouds;
cf.~Section \ref{waves}.
 
In a numerical simulation, Seidel and Suen \cite{SeSu94} have shown
for a spherically symmetric mini-BS, that there exists a
dissipationless relaxation process calling {\em gravitational cooling};
cf.~\cite{Se90,SeSu90,BSS98}.
This mechanism is similar to the violent
relaxation of collisionless stellar systems which settle to a centrally
denser system by sending some of their members to larger radii.
Likewise, a bosonic cloud would gravitationally cool to a
BS by ejecting parts of the scalar matter.
It seems that the influence of the
viscous term in the KG equation (\ref{phi153}) is not efficient enough,
so that radiation of scalar particles is the only viable mechanism.
This was demonstrated numerically
by starting with a spherically symmetric configuration with
$M_{\rm initial} \geq M_{\rm Kaup}$, i.e.~which is more massive than
the Kaup limit, and then monitoring its time evolution
during which the star oscillates and scalar matter is radiated.
Actually, such oscillating and pulsating BS branches were
predicted earlier in our stability analysis via catastrophe theory
\cite{KMS91,SKM92}; cf.~Section \ref{catastrophe}.
Oscillating soliton stars were constructed as well
by using real scalar fields which are periodic in time \cite{SeSu91};
cf.~Section \ref{RSFBS}.
Without imposing spherical symmetry, the emission of
gravitational waves have to be included. In Ref.~\cite{DW96},
it is argued that the purely radial collapse considered in \cite{SeSu94}
is too artifically. A primordial bosonic cloud could have had too
large angular momentum so that the minimal densities required for
BS formation via gravitational cooling are not reached.

The situation of Bose gas clouds, partly even before
Bose-Einstein condensation, was investigated in \cite{IR88} and
\cite{DG98}. Details of \cite{IR88} are exhibited in \cite{Je92}
already, so that we concentrate on the scenario of a
non-isothermal Bose gas cloud
\cite{DG98} in Section \ref{fluid} in which BS formation in different
perfect fluid approximations are outlined.

The influence of a cosmological phase transition on a BS with a field
quantised real scalar is considered in Section \ref{Q-RSF-BS}.

%*********************************************
\subsection{Hot boson stars \label{hot}}

The temperature dependence of the
soliton star of Friedberg et al.~\cite{FLP87b} is investigated in
\cite{SP89}.
In the calculation for the case without gravitational coupling,
i.e.~for a non-topological soliton (NTS), a critical temperature
of $T_c=0.7875 \sigma_0$ is derived and, in the following,
it is assumed that the influence of gravitation does not
change $T_c$ very much. Furthermore, the zero temperature estimates
of \cite{FLP87b} can be applied as well up to a temperature
of $T=0.7 \sigma_0$. Without gravity, the mass of the NTS
vanishes at $T_c$, prohibiting their existence for $T>T_c$;
the same holds for the soliton star.
Below the critical temperature, radius and mass of a soliton star
increase with temperature as one may expect since the kinetic
energy of the particles grows and the potential barrier between true
and false vacuum gets reduced with higher temperatures.
If spontaneous symmetry breaking is restored at $T_{\rm c}$,
the soliton star as well as the NTS vanishes.

The scenario of a mini-BS with temperature is considered
in \cite{BiH00}. Again, a critical temperature is derived numerically
($T_c=0.582 m$): above the bosons are in a gaseous phase,
at $T_c$ there is a second order phase transition, and
below there is a mixed soliton-gas phase. A hot mini-BS
with $m = 1$ GeV/$c^2$ has a radius in the order of magnitude
of cm which is about 13-14 orders of magnitude
larger than in the cold state, cf.~Table IV.
If the scalar field mass is close to the Planck mass,
the condensation begins with a first order phase transition.
For the case $m=M_{\rm Pl}/10$, $T_c=0.00607 m$.

Soliton BHs at finite temperature are mentioned in
Section \ref{SBH}. For real scalar BSs, temperature (or some
kind of motion inside the star as in oscillating BSs) is
decisive for its existence, see Section \ref{RSFBS}.

\subsection{Boson stars in Jordan-Brans-Dicke theory \label{JBD}}

In the so-called
scalar-tensor (ST) theories, a real scalar field replaces Newton's
gravitational constant $G$, the strength of whose coupling to the metric
is given by a function $\varpi(\phi_{\tt JBD})$.
In the simplest scenario,
the Jordan-Brans-Dicke (JBD) theory \cite{BD61}, $\varpi$ is a constant.
GR is attained in the limit $1/\varpi \rightarrow 0$.
To ensure that the weak-field limit of this theory agrees
with current observations, $\varpi$ must exceed 500 at 95\% confidence
limit \cite{W93} from solar system timing experiments, i.e.~experiments
taking place in the current cosmic time. This limit is both stronger and
less model-dependent than bounds from nucleosynthesis \cite{CASAS}.
Scalar tensor theories have regained popularity through inflationary
scenarios based upon them \cite{13-boson}, and because a JBD model with
$\varpi =-1$ is the low-energy limit of superstring theory
\cite{FT85,CFMP85,L85}.
 
In this part, we exhibit the BS theories which contain beside the complex
scalar field of the BS matter, the real scalar field of a JBD type
theory.
We point out the differences in the Lagrangians, but hope also to lighten
the links between the papers. The results lead to observable distinctions
between the models.
In the BS matter part, the authors use in general the potential
$U_{\rm CSW}$ so that a scalar field self-interaction can be taken
into account.
 
The Lagrangian for our system of ST gravity coupled to a
self-interacting, complex scalar field in the (physical) Jordan frame
is
\be
    {\cal L} = \frac{\sqrt{\mid \tilde g \mid}}{2}
  \left [ \phi_{\tt JBD} \tilde{R}
 - \frac{\varpi(\phi_{\tt JBD})}{\phi_{\tt JBD}}
  \tilde{g}^{\mu\nu}  \partial_\mu \phi_{\tt JBD} \partial_\nu \phi_{\tt
JBD}
 + \tilde V(\phi_{\tt JBD})
 +  \tilde{g}^{\mu \nu}
    \partial_{\mu} \Phi^{\ast} \partial_{\nu} \Phi
 - U_{\rm CSW} \right ]  \; . \label{JBDlagr}
\ee
The gravitational scalar is $\phi_{\tt JBD}$ and $\varpi(\phi_{\tt JBD})$
is the Jordan frame coupling of $\phi_{\tt JBD}$ to the matter.
In more general theories, the real scalar $\phi_{\tt JBD}$ possesses
even a potential $\tilde V$.
The complex scalar $\Phi$ has mass $m$ and is
self-interacting through the potential term $U_{\rm CSW}= U_{\rm
CSW}(|\Phi|^2)$.
 
There is an alternative representation of Lagrangian (\ref{JBDlagr}) in
the so-called Einstein frame. The transition to this frame is effected
by the conformal transformation
\be
   \tilde{g}_{\mu \nu} =  e^{2 a(\varphi_{\tt E})} g_{\mu \nu} \; ,
\ee
where
\be
    \phi_{\tt JBD}^{- 1} = \kappa e^{2 a(\varphi_{\tt E})}
\ee
and $a(\varphi_{\tt E})$ is the {\em Wagoner transformation} \cite{Wa70}
from $\phi_{\tt JBD}$ to the gravitational scalar $\varphi_{\tt E}$
in the Einstein frame. The relationship between $\varpi(\phi_{\tt JBD})$
and $a(\varphi_{\tt E})$ is obtained by requiring
\be
    \left ( \frac{\partial a}{\partial \varphi_{\tt E}} \right )^2
  = \frac{1}{2 \varpi + 3} \; .
\ee
Using the potential
$V(\varphi_{\tt E})=e^{4 a(\varphi_{\tt E})}
 \tilde V[\phi_{\tt JBD}(\varphi_{\tt E})]$,
we find the Lagrangian in the Einstein frame
\be
  {\cal L} = \frac{\sqrt{\mid g \mid }}{2\kappa} \biggl [ R
 - 2 g^{\mu \nu} \partial_{\mu} \varphi_{\tt E} \partial_{\nu}
\varphi_{\tt E}
 + V(\varphi_{\tt E}) \biggr ]
 + \frac{\sqrt{\mid g \mid }}{2} e^{2 a(\varphi_{\tt E})}
    \left [ g^{\mu \nu} \partial_{\mu} \Phi^{\ast} \partial_{\nu} \Phi
 - e^{2 a(\varphi_{\tt E})}U_{\rm CSW} \right ]
   \; . \label{EINlagr}
\ee
 
Mathematically, the transition from the  Jordan to the Einstein frame
is a conformal change of metric, cf.~\cite{Mi77c}, and a
field redefinition or ``renormalisation" of the scalar field, the
so-called
Wagoner transformation. The question which  ``frame" describes
the true, physical metric that measures the
distance between spacetime points, is a subtle one which depends on the
coupling to matter, see
the careful analysis of Brans \cite{Br88}. However, the Einstein frame
is sometimes used for calculations of BSs and in certain situations
the results are close enough to GR.   In the following,
we will indicate which frame has been implemented by the authors.
 
Both the coupling functions $\varpi(\phi_{\tt JBD})$
in the Jordan frame and the $a(\varphi_{\tt E})$ in the Einstein frame
are a priori unknown. There are,
however, some theoretical reasons to motivate explicit forms.
Furthermore, once a particular form for the coupling functions
is taken, there are experimental constraints that can be imposed.
 
In Table V, we refer to publications which consider different
gravitational theories and the frame in which the calculations are done;
in some papers, more than one version is investigated.

\begin{center}
\parbox{14cm}{TABLE V.
Publications on complex scalar field BS models in scalar-tensor theory
calculated either in the Jordan or Einstein frame.}
\end{center}
%----------------------table beginning-------------------------
$$\vbox{\offinterlineskip
\hrule
\halign{&\vrule#&\strut\quad\hfil#\quad\hfil\cr
height2pt&\omit&&\omit&&\omit&\cr
&  {\bf Scalar-tensor theory} && {\bf Jordan frame} && {\bf Einstein
frame} &\cr
height2pt&\omit&&\omit&&\omit&\cr
\noalign{\hrule}
height2pt&\omit&&\omit&&\omit&\cr
& Non-minimal-gravity coupling \hfill
  &&  \cite{BG87} \hfill
  &&              \hfill &\cr
height2pt&\omit&&\omit&&\omit&\cr
& JBD, $\varpi=$ const, $V=0$ \hfill
  &&  \cite{TX92,GJ93,TLS98,TSL98,Wh99,WT99,Wh00} \hfill
  &&  \cite{CS98,BSh98,B99} \hfill &\cr
height2pt&\omit&&\omit&&\omit&\cr
& JBD, $\varpi=$ const, $V\neq 0$ \hfill
  &&                    \hfill
  &&  \cite{Y99,FYBT00} \hfill &\cr
height2pt&\omit&&\omit&&\omit&\cr
& ST\phantom{D}, $\varpi=\varpi(\phi_{\tt JBD})$, $a=a(\varphi_{\tt E})$
\hfill
  &&  \cite{T97,TLS98,WT99,Wh00} \hfill
  &&  \cite{CS98} \hfill &\cr
height2pt&\omit&&\omit&&\omit&\cr}
\hrule}$$
%----------------------table end-------------------------

%\begin{itemize}
%\item
 
BS with non-minimal gravity coupling,
$\phi_{\tt JBD} = 1 + 2 \kappa \xi |\Phi|^2$, $\varpi=V=0$:
%\end{itemize}
In 1987, van der Bij and Gleiser  \cite{BG87}
considered a  special JBD case, for which the gravitational scalar is
actually
a function of the BS complex scalar field, i.e.~the BS matter changes its
own
gravitational field. The effective gravitational constant in the coupling
constant
$1/(16\pi G_{\rm eff}) :=\phi_{\tt JBD}/(2\kappa) = 1/(16\pi G) + \xi
|\Phi|^2$ can become
infinite if the free parameter $\xi$ is negative.
The following critical masses and particle numbers arise
\ba
\xi & \rightarrow & +\infty \; , \quad
M_{\rm crit} \approx
  0.73 \sqrt{\phantom{|}\xi\phantom{|}} \, M^2_{\rm Pl}/m \; , \quad
N_{\rm crit} \approx
  0.88 \sqrt{\phantom{|}\xi\phantom{|}} \, M^2_{\rm Pl}/m^2 \; , \quad
 \\
\xi & \rightarrow & -\infty \; , \quad
M_{\rm crit} \approx 0.66 \sqrt{|\xi|} \, M^2_{\rm Pl}/m \; , \quad
N_{\rm crit} \approx 0.72 \sqrt{|\xi|} \, M^2_{\rm Pl}/m^2 \; . \quad
\ea

%\begin{itemize}
%\item
BS in JBD theory and coupling between $\phi_{\tt JBD}$ and $\Phi$,
$\varpi=\kappa f^2/4$, $V=0$, $m^2 \rightarrow m^2 \phi_{\tt JBD}$:
%\end{itemize}
In \cite{TX92}, Tao and Xue applied a JBD theory with positive constant
$\varpi$ including a free parameter $f$. It allowed for a
space-dependent
boson mass, determined by the JBD gravitational scalar, the corresponding
term in the complex scalar field part reads
$m^2 \phi_{\tt JBD} |\Phi|^2$.
After redefinition, $\varpi=G \bar f^2/4$, calculations are done for
$\bar f^2=0,1,2\cdot 10^3$ and a self-interaction constant
$\Lambda=0,2,20,60$ (redefined in our units, see Appendix).
For $\bar f^2=1$, the critical values are
\be
M_{\rm crit} \approx 0.17 \sqrt{\Lambda} \, M^2_{\rm Pl}/m \; , \quad
N_{\rm crit} \approx 0.18 \sqrt{\Lambda} \, M^2_{\rm Pl}/m^2 \; . \quad
\ee

%\begin{itemize}
%\item
BS in JBD theory, $\varpi=6$, $V=0$:
%\end{itemize}
Gundersen and Jensen \cite{GJ93} used a pure JBD theory
with a potential $U_{\rm CSW}$ and
$\varpi=6$ is chosen for the early universe. The critical mass is found
to be
\be
M_{\rm crit} \approx 0.213 \sqrt{\Lambda} \, M^2_{\rm Planck}/m  \; .
\ee
Thus, the deviation of the critical
mass is just a few percent from the GR limit value 0.218.

%\begin{itemize}
%\item
BS in ST theory in Jordan frame,
$\varpi=\varpi(\phi_{\tt JBD})$, $V=0$:
%\end{itemize}
In 1997, Torres \cite{T97} selected three different functional forms
for the JBD coupling function, namely
\ba
1. & \quad & 2 \varpi(\phi_{\tt JBD}) + 3
 = \frac{2 B_1}{|1-\phi_{\tt JBD}/\phi_{\tt JBD,0}|^\alpha} \; ,
  \label{JBD1.} \\
2. & \quad & 2 \varpi(\phi_{\tt JBD}) + 3
 = \varpi_0 \phi_{\tt JBD}^n \; , \\
3. & \quad & \varpi(x) = 0.1x\, ,\; 10x\, ,\; \ln(x)\, ,\; \exp(0.01x) \;
,
\ea
where $B_1,\alpha,n$ are all positive constants,
$\varpi_0$ a constant, and
$\phi_{\tt JBD,0}$ is the asymptotic value of the inverse of Newton's
constant (e.g., the present value). Choice 1.~had already been
investigated in a cosmological context \cite{BP97},
for case 2.~an analytical solution is known \cite{BM94} and a
constraint from nucleosynthesis was determined \cite{T95}.
The intention for explicit functions in situation 3.~was
to understand how the behaviour
of $\varpi$ can be managed within the radius of the star.
Calculations have been done in the Jordan frame for several combinations
of
$B_1=2,5,8$ together with
$\alpha=0.5,1.0,1.5,2.0$ and $\Lambda=0,10,50,100,150,200$
or $\varpi_0=2$ and $n=3$, respectively. Again
the order of magnitude of the BS mass  varies only  about a few percent
 when compared with GR.
The influence of different external asymptotic values of the
gravitational scalar $\phi_{\tt JBD}$ on the BS mass has been
investigated as well with the result that the mass decreases with
increasing time.

%\begin{itemize}
%\item
BS in JBD/ST theory in Einstein frame,
$a=a(\varphi_{\tt E})$, $V=0$:
%\end{itemize}
Comer and Shinkai \cite{CS98} studied BSs in Einstein frame calculations
using two forms for $a(\varphi_{\tt E})$, one being the Brans-Dicke
coupling
\be
  a(\varphi_{\tt E}) =
\frac{\varphi_{\tt E} - \varphi_{\tt E \infty}}{\sqrt{2 \varpi_{\tt JBD}
+ 3}}
 \; ,
\ee
with the constant $\varpi_{\tt JBD}=600$,
and the quadratic coupling
\be
  a(\varphi_{\tt E}) = \frac{\eta}{2}
   \left ( \varphi_{\tt E}^2 - \varphi^2_{\tt E \infty} \right) \; ,
\ee
which is a  particular form considered by Damour and Nordtvedt
\cite{ND93} except for the additive constant; in this cosmological
model there is a constraint of $\eta > 3/8$. The term
$\varphi_{\tt E \infty}$ represents the asymptotic value of the
gravitational
scalar field. Tests in the solar system provide the constraint
$\varpi_{\tt JBD} > 500$; observations of binary
pulsars require $\eta > - 5$ \cite{DE96}. BS calculations are performed
for
$\eta=0.38$ which is close to the limit $3/8$. In the Jordan frame, the
coupling function is
\be
2 \varpi(\phi_{\tt JBD}) + 3 =
  \frac{1}{\eta^2 \varphi^2_{\tt E \infty} - \eta \ln (G \phi_{\tt JBD})}
\; .
\ee
 
In both cases, solution sequences are produced for $\Lambda=0,10,100$
and for zero, one, and two nodes in the complex scalar field solution;
hence,
we find here the first excited states for BSs in scalar tensor theories.
Again,
there is no significant but still observable difference to GR in the
critical mass values. Interesting is that there is a minimal value of the
central $\varphi_{\tt E}(0)$ for which BSs can exist.
Even more, a non-uniqueness is found, i.e.~two different mass values
exist for one $\varphi_{\tt E}(0)$, and an intersection point at which
these two masses are identically. Furthermore, the stability of such BSs
are discussed by using catastrophe theory following
\cite{KMS91,KMS91b,SKM92,KS92,KS93}.
Constructions of BSs in the early universe proved that the BS
mass decreases if we go back in time as Torres \cite{T97} noted before.
In the early universe, only BS states with positive binding energy did
exist,
meaning that no stable BS could have formed. This result was later
doubted by Whinnett \cite{Wh99} as we are going to mention below.

%\begin{itemize}
%\item
BS in JBD theory in Einstein frame,
$\varpi=60, 600$; $V=0$:
%\end{itemize}
Numerical simulations for the dynamical evolution of spherically
symmetric BSs have been carried out by Balakrishna and Shinkai
\cite{BSh98,B99}.
They also investigated the BH formation of a perturbed equilibrium
configuration on an unstable branch. A perturbed unstable BS can also
migrate
to a stable solution by emitting scalar waves; it is shown how an excited
BS state passes into the ground state (similar to the GR case
calculations some years earlier \cite{SeSu90,BSS98}).
Furthermore, the formation
of a stable BS from a Gaussian scalar field packet is demonstrated in
JBD theory, in much the same way as in GR.

%\begin{itemize}
%\item
BS in JBD/ST theory in the Jordan frame,
$\varpi=\varpi(\phi_{\tt JBD})$, $V=0$, gravitational memory effect:
%\end{itemize}
The main focus in \cite{TLS98} was the {\em gravitational memory} effect
of BSs both in JBD theory with $\varpi=400$ and in ST theory with
$2 \varpi(\phi_{\tt JBD}) + 3 = 10/\sqrt{|1-(\phi_{\tt JBD}/\phi_{\tt
JBD,0})}$
both in the Jordan frame; this ST theory is a
special choice of case 1.~in Eq.~(\ref{JBD1.}) \cite{T97}.
 
When the gravitational coupling is evolving, this has important
implications for astrophysical objects \cite{B92}, because it means
that the asymptotic boundary condition for $\phi_{\tt JBD}$ is a
function of epoch affecting the structure of the BS: (a) The star
can adjust its structure in a quasi-stationary manner.
Two sub cases can be distinguished: The adjustment time $t_{\rm adj}$
is smaller than the  evolution time $t_{\rm U}$ of the Universe;
then the BS will always be in equilibrium with the external development.
If $t_{\rm adj}>t_{\rm U}$, the BS will be (quasi-)static, hence,
``conserving''
the value of $G$ at formation time. This effect is called {\em
gravitational
memory}. Actually, a BS which is not completely decoupled from the
cosmological expansion will  develop slowly.
(b) The BS interior produces a feedback on the asymptotic gravitational
constant. If a high density of BSs formed, they might be able
to reduce the gravitational interaction strength in quite a significant
region around themselves.
 
%\subsubsection{Tensor mass}
%\begin{itemize}
%\item
BS in JBD theory in Jordan frame, $\varpi=-1$, $V=0$, tensor mass:
%\end{itemize}
In \cite{Wh99}, Whinnett analyses three different proposed mass functions
for BS systems for the low-energy limit of superstring theory
where $\varpi=-1$ \cite{FT85,CFMP85,L85}. These are the Schwarzschild
mass corresponding to the ADM mass in the Jordan frame, the Keplerian
mass in the Jordan frame and the Keplerian mass in the Einstein frame
which he calls {\em tensor mass}; cf.~with the earlier investigation in
\cite{Le74}.
In his Figure 1, the results for the mass definitions are compared with
the results for the particle number in the Jordan frame.
The mass definitions differ significantly leading to contrary physical
interpretations for stability. The Keplerian mass in the Jordan frame
yields positive binding energy for all values of central density,
suggesting that every solution is generically unstable.
The Schwarzschild mass in the Jordan frame instead leads to negative
binding energies for every value of central density, suggesting that
every
solution is potentially stable. However, the tensor mass peaks
at the same location as the rest mass, an important property in
the application of catastrophe theory to analyse stability
properties \cite{KMS91}. Therefore, Whinnett adopts the
tensor mass as the {\it real mass of the star}.
We stress the interesting situation that whereas the
(Keplerian) mass is calculated in the Einstein frame, the
particle number (or rest mass) is evaluated in the Jordan frame.
Finally, we remark that, in this paper, the dilaton field $\varphi$
is used, i.e.~$\phi_{\tt JBD}=\exp (-\varphi)$.
 
A comparison of the critical masses shows that for the choice $\varpi=6$
of Gunderson and Jensen, the mass values
are about 7\% different from the results of the tensor mass \cite{GJ93}.
Already for $\varpi=500$ the difference between the masses is less
than $10^{-3}$, i.e.~negligible.
 
Whinnett \cite{Wh99} proved also that the tensor mass and the particle
number peak at the same value of the central density, which,
in GR, is known due to Jetzer \cite{Je90}. In the same year,
Yazadjiev \cite{Y99} gave a similar analytical proof for more general
ST theories with a term $F(\phi_{\tt JBD}) \tilde R$,
$\varpi(\phi_{\tt JBD})=H(\phi_{\tt JBD}) \phi_{\tt JBD}$, and
$V(\phi_{\tt JBD}) \neq 0$ where $F,H,V$ are general functions.

%\begin{itemize}
%\item
BS in JBD theory in Jordan frame, $\varpi=400$, $V=0$:
%\end{itemize}
More physical investigations for such BSs at different cosmic epochs
are carried out in \cite{TSL98}. It results that BSs
can be stable at any time of cosmic history and that equilibrium stars
are
denser in the past applying the tensor mass. It is shown that the radius
corresponding to the critical BS mass remains roughly the same during
cosmological evolution. Several new physical features are displayed:
the mass-radius relation, the behaviour of the difference between the
central and asymptotic $\phi_{\tt JBD}$ value, the dependence of the
binding energy on the coupling constant $\varpi$.

%\begin{itemize}
%\item
Charged BS in JBD/ST theory in the Jordan frame, $V=0$:
%\end{itemize}
In 1999, Whinnett and Torres \cite{WT99} investigated the influence of
an additional $U(1)$ charge for such BSs with constant coupling
function $\varpi$ and a special choice of power-law ST theory.
Solutions are calculated for
the following parameter combinations (always $\phi_{\tt JBD,\infty }=1$):
For $\Lambda=0$ $\varpi=-1,1,500, 16 \phi_{\tt JBD}^4/3$,
for $\Lambda \rightarrow \infty$ $\varpi=1$.
It is found that there is a maximal charge per mass ratio,
$q=e M_{\rm Pl}/m$, above which weak field BS solutions are not stable.
For $q=0$, the mass evolution for four different stable BSs is
shown for $\varpi=-1$ and $N=$constant if $\phi_{\tt JBD,\infty }$
changes
from 1 to 6 (Fig.~8 therein); for different constant charges $q$
the same is done in their Fig.~9. For BSs with constant central density
it is
demonstrated that the BS mass increases under certain conditions
for the BS solutions. Additionally, the effect of spontaneous
scalarization is investigated  \cite{Wh00}.

%\begin{itemize}
%\item
BS in JBD/ST theory in Jordan frame, $V=0$:
%\end{itemize}
In \cite{Wh00}, Whinnett considered the appearance of spontaneous
scalarization.
This phenomenon can also develop in neutron stars
as shown in \cite{DE93} in the Einstein frame.
It means that for a star below its critical mass,
the scalar tensor field $\phi_{\tt E}$ is nearly constant throughout
the star but, for more massive stars, $\phi_{\tt E}$ has a large
spatial variation. Besides BSs with $\varpi=1,10,500,3300$,
the function $a(\phi_{\tt E})=-k \phi_{\tt E}^2$ is investigated
in the Einstein frame which is equivalent to the Jordan frame coupling
\be
2 \varpi(\phi_{\tt JBD}) + 3
 = \frac{1}{2k \log \phi_{\tt JBD}} \; .
\label{Wh2}
\ee
The third choice is
\be
\varpi(\phi_{\tt JBD}) = \frac{\phi_{\tt JBD}}{8\xi (\phi_{\tt JBD}-1)}
\; ,
\label{Wh3}
\ee
where $k=1,3$ and $\xi=1,2$ are the constants considered
and $\phi_{\tt JBD}\ge 1$ in the third choice. For all three cases,
the complex scalar field self-interaction constant is set to 0;
additionally, for the JBD case
and the coupling function (\ref{Wh2}), $\Lambda \rightarrow \infty$.
The calculations are carried out in the Jordan frame.
Whinnett concludes that there is no spontaneous scalarization if
$\Lambda =0$, but it appears in the situation with a strong
self-interaction,
and so, ultimately depending on the compactness of the star.

%\begin{itemize}
%\item
BS in JBD theory in Einstein frame, $\varpi=0$, $V=V(\phi_{\tt E})$:
%\end{itemize}
Fiziev et al.~\cite{FYBT00} examined the influence of different
potentials
$V$ giving a mass to the dilaton field $\phi_{\tt E}$, cf.~\cite{FI92}.
The chosen potentials are:
(a) $V=(3m_{\rm dil}^2/2) [1-\exp(2\phi_{\tt E}/\sqrt{3})]^2$
which corresponds to $V=(3m_{\rm dil}^2/2) (\phi_{\tt JBD}-1)^2$
in Jordan frame,
(b) $V=2m_{\rm dil}^2\phi_{\tt E}^2$,
(c) $V=2m_{\rm dil}^2 \sin^2(\phi_{\tt E})$,
(d) $V=4m_{\rm dil}^2 [1-\cos(\phi_{\tt E})]$,
(e) $V=2m_{\rm dil}^2 [1-\exp (-\phi_{\tt E}^2)]$.
The parameter $\gamma=m_{\rm dil}/m$ is introduced and all calculations
exhibited in the paper are performed for potential (a) and $\Lambda=20$
in our conventions;
it is stated that potentials (b)-(e) does not change the results
noticeable.
It is discovered that for large values of $\gamma \sim 10$ the GR
solutions are found. An increasing $\gamma$ means an
increasing dilaton mass with respect to the boson mass leading to a
smaller dilaton range and, hence, suppressing the dilaton field.
Additionally, the stability analysis for the BS states is performed using
catastrophe theory.

\subsection{Boson halo \label{halo}}
 
In 1994-96, two different models for fitting rotation curve data
of spiral galaxies were proposed using BS-like objects
\cite{Sin94,JS94,Sch94,Sch95,LK96,Sch97,Sch98,Sch98b,Sch99}.
In \cite{Sin94}, solutions with $n=5$ and $n=6$ of the
Newtonian BS system (\ref{Newt1}), (\ref{Newt2}) were applied to
fit the ripples of spiral galaxy NGC2998 without the contribution of
visible matter. The influence of visible matter was included in
\cite{JS94}
for NGC2998 and NGC801 by assuming some functional matter distributions
for the bulge and the disk. In \cite{LK96}, the rotation velocities
for a  BS with $\Lambda=0$ and $\Lambda=300$
for $n=9$ are shown without fitting rotation curve data.
In all three papers, just the Newtonian rotation
velocity formula $v^2(r)=M(r)/r$ was applied, i.e.~the non-negligible
influence of the radial pressure of the anisotropic BS matter was
ignored; the correct formula \cite{Sch98} reads
\be
v^2(r)= \frac{M(r)}{r} + \frac{\kappa}{2} p_r(r) r^2 e^{\lambda(r)
+\nu(r) }
\, .
\ee
 
In Refs.~\cite{Sch94,Sch95}, Newtonian solutions for the massless case
of the Lagrangian (\ref{lagrBS}) were investigated.
Later these solutions were designated as
{\em boson halos} \cite{Sch98}. It was shown \cite{Sch94}
that the boson halo mass increases linearly with radius so
that the rotation velocity is constant with distance; both oscillate
slightly, e.g.~the rotation velocity around a constant value as
\be
v^2(r)=|\Phi(0)|^2 \left [ 1 - \frac{\sin(2\omega r)}{2\omega r} \right ]
\; .
\ee
In \cite{Sch95}, massless
self-interacting complex scalar fields show a similar behaviour.
However, for too strong gravitational fields, independent on
the self-interaction constant, a singularity forms at the centre;
cf.~\cite{M99} where also the case of a massless conformally coupled
complex scalar field with $U= R |\Phi|^2/6$
leads to a linearly increasing mass.
 
A more detailed investigation of the boson halos included the influence
of the visible HI gas and star matter \cite{Sch97,Sch98,Sch98b}.
The spiral and dwarf galaxies fitted in \cite{Sch98} belong to a
sample of 11 galaxies
of different Hubble types and absolute magnitudes which fulfil
several strong requirements \cite{BBS91}. All rotation curve data are
measured in the 21-cm line of neutral hydrogen, so that the gas
distribution extends far beyond the optical disc (at least 8 scale
lengths)
and the necessity of dark matter is becoming obvious. Isolation of
galaxies are another constraint so that perturbative effects of
nearby situated galaxies are negligible.
{}For dwarf galaxies, the
dark matter density near the centre is almost constant, i.e.~dark
matter has a core in these galaxies. The Newtonian solutions of
\cite{Sch98} for a massless complex scalar field reveal a constant
density near the centre as well, and so, good fits for dwarf galaxies
could
be obtained. In all fits for dwarf galaxies, it is recognisable that the
dark matter halo dominates the luminous parts of the galaxies. Especially
the maximum (and the following drop) of the DDO154 data can be matched
perfectly by the boson halo; i.e.~decreasing
rotation curve data, near the end of observational resolution,
can be explained by using a dominating dark matter component.
The sample of rotation curve data of spiral galaxies are fitted as well
as the universal rotation curves of \cite{PSS96}.
Furthermore, oscillating behaviour within the optical
rotation curves of some galaxies can be found, e.g.~in NGC2998,
which can be fitted by a dominating boson halo part.
The general relativistic configurations for the boson halo
reveal very large gravitational redshifts (calculated up to a value of 20).
In \cite{BSc98}, it is shown that boson halos are stable.
 
Recently, it has been shown that BS solutions with a soliton type
potential,
$U=m^2|\Phi|^2 (1-\lambda_6 |\Phi|^4)$, can be used as well
to fit rotation curve data of spiral and dwarf galaxies
\cite{MS02b,MSP02}.
Further investigations on boson halos with a massive complex
scalar field derive a scalar field mass of about $10^{-23}$ eV/$c^2$,
if a self-interaction is included the scalar field weighs about 1
eV/$c^2$, cf.
\cite{ALS01,ALS02}. Cosmological consequences are considered
in \cite{HBG00,SW00}. A prediction on core radii from a formation
scenario
is given in \cite{RT00}.
 
In a different approach \cite{DR93,DRA95}, the dark matter part
consists of a gas of massive bosonic particles, and
a total of 36 galaxies were fitted including also a contribution for the
baryonic matter. The best fits for all galaxies is achieved for
a mass of about 60-70 eV/$c^2$ of the bosonic particles.

\subsection{Classical Klein-Gordon particle}
 
Conventionally, a BS turns out to be an object of a huge number $N$ of
particles, but there are also some attempts to construct a BS with just
one particle \cite{FM68,Ro94}.
Feinblum and McKinley \cite{FM68}, shortly before the work of Kaup,
investigated the mini-BS case
with two constraints, namely the normalisation  of both the particle
number $N=\omega$ and of the mass $M=\omega^2/m$. This leads to exact
{\em one} ground state particle solution, possessing
a vanishing binding energy by construction. The authors had
numerical problems to find a normalised ground state;
this may be due to the chosen mass value, $m^2=0.1$, which seems not to
be
the appropriate value of the single ground state,
as Kaup has pointed out \cite{K68}. We may add
that, in flat spacetime, the KG equation does not permit a one
particle interpretation, as is well-known. In geometrodynamics,
there are different approaches to describe elementary particles
\cite{W61,Mi77a,Mi77b,Mi80,Mi81}.
 
In 1994, Rosen \cite{Ro94} constructed particle-type solutions
for the linear KG equation which he called {\em Klein-Gordon particles}.
The first constraint is that the total mass is equal to the
particle mass, i.e.~$M=m$. The second constraint is rather ad-hoc,
namely $m=\omega$. Furthermore, it is assumed that these
particles have a clear surface where the scalar field is zero and
the vacuum Schwarzschild metric starts to be present, similarly as
for neutron stars. The mass of the particles are in the order of the
Planck mass.
In a follow-up paper \cite{Ro94}, Rosen constructs {\em Proca particles}
with the same conditions as above. Due to an oscillating electric
potential, necessarily also an oscillating magnetic vector
potential is needed for this construction.
In Ref.~\cite{M99}, it was noted that the radial pressure for the
KG and the Proca particle does not vanish at the surface,
thus these solutions are unstable.

\subsection{$D$-dimensional boson stars \label{2+1}}
 
Whereas in four dimensions only numerical BS solutions are known,
a complex scalar with large self-interaction constant $\Lambda \gg 1$
allows analytical solutions in (2+1) dimensions.
Spherical symmetry reduces in (2+1)
dimensions to circular symmetry. Such circular BSs in a spacetime
with negative cosmological constant exist and can be stable
as long as the absolute value of the negative cosmological constant
is lower than some critical one, otherwise all BSs have positive
binding energy and are not energetically favourable.
This result was derived for rotating BSs as well,
and for non-rotating boson-fermion stars \cite{SaSh98,SaSh98b,SaSh00}.
Furthermore, the rotating BS shows a vacuum hole at its centre,
similar to the (3+1) dimensional counterpart, cf.~Section \ref{rotaBS}.
An analytical solution for a Newtonian (3+1)-dimensional BS in the limit
$\Lambda \gg 1$ is determined in \cite{LK96}, cf.~Section \ref{NewtBS}.
 
Recently, mini-BS solutions in three, four, or five dimensions with
an asymptotic spacetime with negative cosmological constant
has been constructed in Ref.~\cite{AR02}. The effect of the
negative cosmological constant on the BS interior is an
additional attractive interaction. Therefore, the mini-BS mass
decreases with decreasing negative cosmological constant.
Whereas a numerical investigation of the radial perturbations
shows that stable mini-BSs exist, from the Figures, it follows
that all BSs have positive binding energy and, hence, are
potentially unstable.

\subsection{Gauge boson solitons}
 
There exists a huge amount of papers on self-gravitating
$SU(n)$-solutions, i.e.~for Einstein-Yang-Mills
theories, and combinations with other matter fields
\cite{BM88,BFM92,BFM94,VG99}.
These globally regular solutions can be regarded as
{\em gauge BSs}, cf.~\cite{Sch93}, except that they are unstable
\cite{SZ90,BS96}. Abelian fields occur in the
{\em gravitational-electromagnetic entities} ({\em geons}) of Wheeler
\cite{Wh55}; cf.~\cite{AB96}.
If the gravitational influence is weakened so much that
even Newtonian gravity can be neglected, non-topological or topological
solitons are obtained \cite{FLS76,Le81}, cf.~\cite{Ac79}, or
for cylindrical symmetry, --- in a way --- {\em thread `BSs'}
can be found \cite{OS96,OS97}.

%********************************
\subsection{Axidilatons from effective string models\label{axisect}}
 
Commonly, in four-dimensional {\em effective} string theories
\cite{DP94},
the tensor field $g_{\mu\nu}$ of gravity is accompanied by
several scalar fields.
Besides the familiar dilaton $\varphi$,
the `universal' {\em axion} $a $, a pseudo-scalar
potential for the Kalb-Ramond (KR) three-form
$H:=e^{\varphi/f_\varphi}\, ^*da $ arises.
Through  spontaneous compactification from ten dimensions onto an
isotropic six torus of radius $e^\beta$, a further modulus field $\beta$
emerges.
 
In the string frame
$\widetilde g_{\mu\nu} = e^{-\varphi/f_\varphi}\,g_{\mu\nu}$,
the {\em effective} string Lagrangian reads
\be
{\cal L}_{\rm eff} = \sqrt{\mid \widetilde g\mid }e^{-\varphi/f_\varphi}
\left[{\widetilde R\over{2\kappa}}+ \widetilde g^{\mu \nu}
\left(\partial_\mu \varphi\partial_\nu \varphi -
6 \partial_\mu \beta\partial_\nu \beta -
\frac{1}{2}
e^{  2\varphi/f_\varphi}\partial_\mu a \partial_\nu a \right)
-\frac{1}{2} e^{\varphi/f_\varphi} \widetilde U(\varphi,a  )  \right]
\, ,  \label{efflag}
\ee
cf.~Eq.~(11) of Dereli et al.~\cite{DOT95}.
Via conformal change \cite{Mi77c} of the metric
$g_{\mu \nu} \to {\widetilde g}_{\mu \nu}=
\exp(\varphi/\sqrt{2\kappa}) g_{\mu \nu}$, we can go over to the Einstein
frame.
Thereby, the kinetic dilaton term changes
sign and allows us to formally combine \cite{Se93}
the axion and the dilaton into a {\em single complex} scalar field
\be
\Phi:=  a  +i f_\varphi e^{-\varphi/f_\varphi}\,,
\ee
the {\em axidilaton}.
For this {\em complex} scalar field the Lagrangian reads
\be
{\cal L}_{\rm eff} = \sqrt{\mid g\mid }
\left[ {R\over{2\kappa}} - g^{\mu \nu}
\left(\frac{1}{2} e^{2\varphi/f_\varphi} \partial_\mu
\Phi\partial_\nu\Phi^*
+ 6 \partial_\mu \beta\partial_\nu \beta \right )
- \frac{1}{2}e^{2\varphi/f_\varphi} \widetilde U(\varphi,a  )  \right]
\, .
\ee
The emerging self-interaction has, in lowest order, a mass term
\be\label{axipot}
U(\varphi,a ):= e^{2\varphi/f_\varphi}\widetilde U(\varphi,a )
  \simeq
 m_a ^2 \mid \Phi\mid^2 \, ,
\ee
which exhibits residual $U(1)$ symmetry, resembling symmetry
restoration. Thus, the
$U(1)$ sector of this effective string model is invariant under the
global
Noether symmetry $\Phi \rightarrow e^{i\alpha}\Phi$ which allows us to
establish again the {\em global stability} \cite{KMS91,KMS91b} of the
star.
 
All the occurring masses $m$ and decay
constants $f$ of the bosonic particles are related via
$m_\varphi f_\varphi \simeq m_a  f_a  \simeq m_\pi f_\pi\simeq
10^{16}$ (eV)$^2$ to those of the pion.
Whereas the mass of the {\em dilaton} $\varphi$ is conventionally
related \cite{Da99} to the supersymmetry breaking scale
$m_{\rm SUSY}$ by $m_\varphi \simeq 10^{-3}(m_{\rm SUSY}/$TeV/$c^2)$
eV/$c^2$, this is not the only possibility.
In the context of {\em string cosmology} \cite{GV99},  massless
axions are able to seed the observed anisotropy of the Cosmic Microwave
Background (CMB). The same holds for the KR axion, provided it
lies in an ultra-low mass window.
Besides the full dilaton interaction
for the bosons as mentioned above, we could consider a
very light dilaton $\varphi$ which is
{\em stabilised} \cite{Di97,Di97b} through the axion.
 
Instead of combining the dilaton with the axion, one could
also take recourse to the other moduli field and consider, e.g.,
the complex {\em K\"ahler form field}
$\hat\Phi:= \varphi + i \sqrt{12} \beta$ akin to T-duality \cite{Du95},
without including the axion.
 
%********************************
 
{\em Are MACHOs axidilaton stars?}
In Ref.~\cite{MS00,MS01}, we have
proposed that the weakly interacting dilatons and axions,
after a partial Bose-Einstein condensation to an {\em axidilaton star}
(ADS), account also
for some fraction of the MACHOs ({\em MA}ssive {\em C}ompact {\em H}alo
{\em O}bjects) which are dark matter candidates, almost certainly
detected
by {\em gravitational microlensing} \cite{Su99}, though still with
statistical ambiguities \cite{GK02}.
Since the total mass of an ADS is not well constrained, one could
phenomenologically turn the argument around:
By identifying the MACHOs of gravitational mass
of about 0.6 $M_\odot$ (as given by microlensing observations)
with the critical mass of an ADS, we are
essentially ``weighing", via $M_{\rm Kaup}/N_{\rm crit} \cong m$,
the {\em constituent} mass to be of the order
$m_a \simeq 10^{-10}$ eV/$c^2$. It is
encouraging to note that such an ultra-low mass
is perfectly compatible with the constraints on the
mass range of the KR axion
seeding the large-scale CMB anisotropy, cf.~the recent results of
Gasperini and Veneziano \cite{GV99} within low-energy
{\em string cosmology}.
 
If ADSs during their evolution would accumulate a larger mass,
they can start to oscillate \cite{KMS91} and thereby
get rid of some excess mass due to `gravitational cooling'
\cite{SeSu94}. Repeated accretion leads ultimately
to a BH in the upper range of the
MACHO mass. These axion induced BHs \cite{Se92,GGK95}
do not carry scalar hair, which
could serve to distinguish them from primordial BHs, but could
have some remnant of P-violation or even {\em ``axion hair"}
\cite{DKO92},
if $a $ is interpreted as a superpotential for the KR axial torsion.
In the pre-big bang scenarios, the dilaton
would produce a detectable gravitational wave background \cite{CLLW98}.
 
Therefore, if such string-inspired scalar fields would exist in
Nature, axions could not only solve the {\em non-baryonic} dark matter
problem \cite{TT99}, but their gravitationally confined
{\em axidilaton stars} would also represent the
MACHOs which presumably account for about 20\%
of the halo dark matter in our Galaxy.

%*************************************
\subsection{Quantised complex scalar field boson star\label{Q-CSF-BS}}
 
Field quantisation of a complex scalar has been studied in \cite{Gr90b}
in the context of BS formation. In Ref.~\cite{HKL02}, a BS with
a field quantised complex scalar has been introduced.
 
As is well-known, the complex scalar $\Phi$ can be regarded as
two real fields, namely $\phi_1=(\Phi +\Phi^*)/\sqrt{2}$ and
$\phi_2=i (\Phi -\Phi^*)/\sqrt{2}$.
Then, field quantisation results in the operator expansion
\be
\phi_j  =  \sum_{nla} b_{nla}^{(j)\dagger} R_l^{(j)n}(r)
 Y^{|a|}_l(\theta, \, \varphi)
  e^{-i \omega_{nl}^{(j)} t} + b_{nla}^{(j)} R_l^{(j)n}(r)
 [Y^{|a|}_l(\theta, \, \varphi)]^*
  e^{+i \omega_{nl}^{(j)} t}\,,
\ee
where $j=1,2$.
For a {\em bound state}, $R_l^{(j)n}(r)$ are radial distributions as
generalisations of the wave function of the hydrogen atom,
$Y^{|a|}_l(\theta, \, \varphi)=(1/\sqrt{4\pi})
P^{|a|}_l(\cos\theta)\, e^{-i|a|\varphi}$ the spherical harmonics
given in terms of the
normalised Legendre polynomials, and $|a|\leq l$ are the quantum numbers
of {\em azimuthal} and {\em angular} momentum.
The non-vanishing  commutation relations for the bosonic creation and
annihilation operators are
\be
[b_{nla}^{(j)},b _{n'l'a'}^{(j)\dagger}] =
  \delta_{nn'} \delta_{ll'} \delta_{aa'} \, .
\ee
In general, there are two number operators
$N^{(j)}:=b^{(j)\dagger}b^{(j)}$.
{}For the ground state of a cold BS, if $N_1=N_2$, we have
$|N\rangle=|N,0,0,\ldots\rangle:=\prod_1^N b_{100}^{(1)\dagger}
\vert0\rangle
  =\prod_1^N b_{100}^{(2)\dagger} \vert0\rangle$,
where $\vert0\rangle$ is a vacuum state in the curved spacetime
`background'; cf.~Section \ref{Q-RSF-BS}.
 
In Ref.~\cite{HKL02} an attempt was made to
motivate {\em symmetry breaking}
in a model with $|\Phi|^4$ self-interaction and self-gravity.
Three different BS models lead to three different
effective coupling constants. The BS with a not quantised
complex scalar \cite{CSW86} has simply $\Lambda/2$.
The BS model with a field quantised real scalar produces $3\Lambda/4$;
and the BS model with a field quantised complex scalar has $5\Lambda/8$.
All three models results in the limit $\Lambda \gg 1$
to small changes in the critical mass
$M_{crit} \sim K \sqrt{\Lambda} M_{\rm Pl}^2/m$, where
$K=0.22$, 0.225, and 0.223, respectively.

%**************************************
\section{How to detect complex scalar field boson stars}
 
In this section, we speculate on observational consequences
for the spherically symmetric BS
model given by (\ref{lagrBS}), if not mentioned differently.
In several cases, we shall recognise the important influence of the
scalar field potential.
Under our assumption that the complex
scalar field has no interaction with baryonic matter other
than gravitationally, the BS is transparent, i.e.~baryonic
matter can accumulate up to the centre of the BS
or photons and particles can move through the BS interior, respectively.
 
To detect a BS depends on the physical situation; there is a
corresponding wave-band where the BS might be very luminous.
In the following, we shall discuss several of such scenarios.

\subsection{Rotation curves\label{curves}}
 
It might be possible to detect a BS in X-rays. Imagine a
massive BS, say of $10^6 M_{\odot}$, for which it is likely that
an accretion disk
forms and since its exterior solution is asymptotically Schwarzschild,
nearly beyond its effective radius the BS looks similar
to an {\em A}ctive {\em G}alatic {\em N}ucleus (AGN) where usually a
black hole (BH) at the centre is assumed.
In X-rays \cite{SL97,SL98,Sch02}, one can probe close to the
Schwarzschild radius or, as we propose, even inside the BS;
Iwasawa et al.~\cite{Iw96} claimed that by using data from the
satellite ASCA they have
probed to within 1.5 Schwarzschild radii; cf.~\cite{LFB99,MKM99,YL99}.
Because this is inside the
static limit for a non-rotating BH, they conclude that a Kerr
geometry is required. A BS configuration could
still be an alternative explanation, giving a non-singular solution where
emission can occur even from the centre.
 
In order to differentiate between a BS and a BH, it is interesting to
calculate the tangential velocities of  baryonic matter that may rotate
around the centre of a BS. For the static spherically
symmetric metric (\ref{Schw}), geodesics of a collisionless circular
orbit obey
\be
v_\varphi^2 =\frac{1}{2} r \nu' e^{\nu } =
\frac{1}{2} e^{\nu } (e^{\lambda}-1) +
        \frac{\kappa}{2} p_r r^2 e^{\lambda +\nu } \simeq
        \frac{M(r)}{r} +  \frac{\kappa}{2} p_r r^2 e^{\lambda +\nu }
        \; ,
\ee
which reduces for a weak gravitational field into the Newtonian form
$v_{\varphi,{\rm Newt}}^2 = M(r)/r$ if $p_r=0$.
Rotation curves for the cases $\Lambda =0$, 10 and 300 were calculated
for a critical mass BS in \cite{SL97,SL98}.
The curves increase from zero at the centre up to a maximum followed
by a Keplerian decrease. The maximal rotation velocities reaches
still inside the BS more than one-third of the velocity of light;
similar results attain for the potentials
$U_{\rm CG},U_{\rm SG},U_{\rm L}$ (Fig.~8 of \cite{ST00}).
For a BS in  Newtonian approximation, the maximal circular velocity
can be just about 2000 km/s and this at more than 50 times
of its Schwarzschild radius, but still inside the BS.
One can conclude that large rotation
velocities $v_{\varphi}$ are not necessarily signatures of BHs.
Baryonic matter rotating with the maximal velocity of about
$c/3$ possesses an impressive kinetic energy of up to
6\% of its rest mass.  If one supposes that each year a
mass of 1$M_\odot $ transfers this amount of kinetic energy into
radiation, a BS would have a luminosity of 10$^{44}$erg/s.

\subsection{Gravitational redshift\label{redshift}}
 
If baryonic matter inside a BS emits or absorbs
radiation within its gravitational potential, the spectral feature
is redshifted \cite{SL97,SL98,Sch02}.
The gravitational redshift $z_{\rm g}$ within a static background
is given by
$1+z_{\rm g} := \sqrt{e^{\nu (R_{{\rm ext}})}/e^{\nu (R_{{\rm int}})}}$,
see e.g.~Ref.~\cite{N93},
where emitter and receiver are located at $R_{{\rm int}}$ and
$R_{{\rm ext}}$, respectively.  The maximal possible redshift for a given
configuration is obtained if the emitter is exactly at the centre
$R_{{\rm int}}=0$. The receiver is always practically at infinity,
hence $\exp \left[\nu (R_{{\rm ext}}) \right]=1$.
For all other redshifts in between, the gravitational
redshift function is defined as
\be
1+z_{{\rm g}}(x) = \exp \left(-\frac{\nu (x)}{2}\right)  \, .
\label{zfunc}
\ee
Table 1 in \cite{SL97} gives the results for the critical mass BS
for different $\Lambda $. The redshift  lies between 0.0657
for $\Lambda=-20$, 0.4565 for mini-BS with $\Lambda=0$, and 0.6873
for the limit case $\Lambda \gg 1$. With increasing self-interaction
coupling constant $\Lambda $ also the maximal redshift  grows.
Unstable BSs presumably may have arbitrary gravitational redshift.
In Ref.~\cite{ST00}, for the $\cosh$-Gordon potential $U_{\rm CG}$, the
maximal redshift is found to be 0.46,
for the $U(1)$-Liouville potential $U_{\rm L}$, it is
$z_{\rm max}=0.49$. For a neutron star, Markov \cite{Ma64}
derived $z_{\rm max}=0.49$.
In general, observed redshift values would consist of a combination of
cosmological and gravitational redshifts given by
$(1+z)=(1+z_{\rm c})(1+z_{\rm g})$ due to the addition formula of
relativistic velocities.

%****************************
\subsection{Gravitational lensing\label{lensing}}
 
In this section, the results of gravitational lensing of a {\em
transparent}
spherically symmetric mini-BS are shown \cite{DS00,SD02}.
In contrast to the last sections, it is assumed that the BS interior
is empty of baryonic matter, so that deflected
photons can travel freely through the BS. The deflection angle
\cite{Wa84}
is then given by
\be
\hat{\alpha}(r_0) = 2 \int\limits_{r_0}^\infty
\frac{be^{\lambda /2}}{r\sqrt{r^2 e^{-\nu} - b^2}} dr  - \pi \; ,
\ee
where $b$ is the impact parameter
and $r_0$ denotes the closest distance between a light ray and the centre
of the BS.
The lens equation {\em for small deflection angles} can be expressed
\cite{NB96} as
\be
\sin\left(\vartheta-\beta\right)
 = \frac{D_{\rm ls}}{D_{\rm os}} \sin\hat{\alpha} \, ,
 \label{lens1}
\ee
where $D_{\rm ls}$ and $D_{\rm os}$ are the distances from the lens
(deflector) to the source and from the observer to the source,
respectively.
The true angular position of the source is denoted by $\beta$,
whereas $\vartheta$ stands for the image positions.
One usually defines the reduced deflection angle to be
$
\alpha \equiv \vartheta - \beta = \sin^{-1}\left(D_{\rm ls}
\sin\hat{\alpha} /D_{\rm os} \right) \, .
$
However, equation (\ref{lens1}) relies on substitution of the
distance from the source to the point of minimal approach by the
distance from the lens to the source $D_{\rm ls}$, see \cite{NB96}.
{\em For large deflection angles} the distance $D_{\rm ls}$
cannot be considered a constant but it is a function of the deflection
angle
so that the form of the lens equation changes \cite{DS00} into
\begin{equation}
\sin{\alpha} = \frac{D_{\rm ls}}{D_{\rm os}} \cos{\vartheta}
\cos \left[ \mbox{arsin}
\left ( \frac{D_{\rm os}}{D_{\rm ls}} \sin (\vartheta - \alpha )
  \right) \right]
\left[\tan{\vartheta} + \tan(\hat{\alpha} - \vartheta)
\right] \, .  \label{lens2}
\end{equation}
The reduced deflection angle can be kept defined as
$\alpha \equiv \vartheta - \beta$.
 
Numerical computations of the reduced deflection angle for different
BS potentials $U_{\rm K}, U_{\rm CG} , U_{\rm SG} , U_{\rm L} $
were performed by assuming that the BS lens is half-way between the
observer
and the source, i.e.~$D_{\rm ls}/D_{\rm os}=1/2$
(for details, see \cite{DS00,ST00}).
Observable differences are found depending on the choice of the
self-interaction. In the case
of a simple mass term  $U_{\rm K}$, the largest possible value of
$\alpha$ is
23.03 degrees with an image at about $\vartheta = n \times 2.88$
arcsec where $n=n(D_{\rm ol},\omega)$ is the distance factor which is a
function of the distance from the observer to the lens and the
scalar field frequency the inverse of which is associated with the
BS radius. For non-relativistic BS approximations, smaller angles will
occur.
The angle $\vartheta$ of the image position can have very
different orders of magnitude, depending on $n$.
For example, $n=1$ fixes $\vartheta$ to be measured in arcsec.
Under the assumption that the BS mass is 10$^{10}$ $M_{\odot}$,
the distance $D_{\rm ol}$ is about 100 pc.
If the distance factor is $n=10^{-3}$,
then $\vartheta $ is measured in milli-arcsec and the
BS-lens is at about 100 kpc from the observer.

\begin{figure}[th]
\centering
\leavevmode\epsfysize=7cm \epsfbox{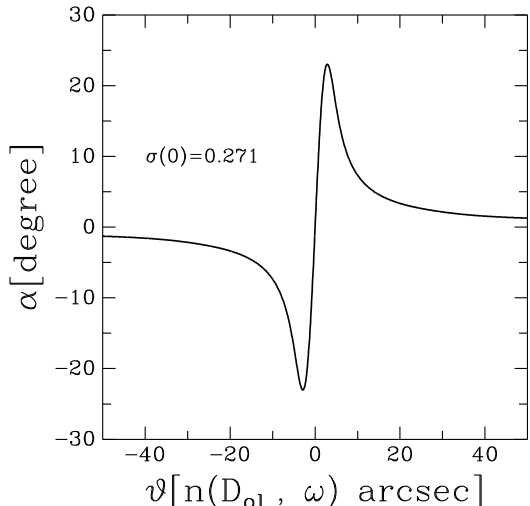}\\
\caption[fig3]{Reduced deflection angle for a mini-BS of critical mass.
The angle $\vartheta$ for the image positions is given in units of arcsec
times the distance function $n$ \cite{DS00}.\label{fig3}}
\end{figure}

In Ref.~\cite{DS00}, it is summarised that the BS has all qualitative
features of a non-singular spherically symmetric transparent lens,
cf.~Fig.~\ref{fig3}, the lens curve for a mini-BS of critical mass.
Three images can be observed, two of them
being inside the Einstein radius and one outside. An Einstein ring
with infinite tangential magnification, the so-called tangential
critical curve, is found, and also a radial critical curve for
which two internal images merge.
The appearance of the radial critical curve distinguishes BSs
from other extended and non-transparent lenses \cite{SEF92}. For a BH
or a neutron star, the radial critical curve does not exist
because it is inside the event horizon or the star, respectively. Two
bright images near the centre of the BS and the third image at
some large distance from the centre are found.
If an extended transparent source like the BS is considered, one can
imagine how to detect the BS. In such a case, two images, one radially
and the other tangentially elongated and both close to each other,
might be observed. By looking along the
line defined by these two images, the third one can be detected at
a large distance.
 
For the Cosh-Gordon potential $U_{\rm CG}$ \cite{ST00}, the maximal
reduced deflection
angle is 23.229 degrees and for $U_{\rm L}$ an even
larger deviation, namely 24.391 degrees, with an image at about the
same place is found. Differences among all four mentioned potentials
are up to 4\%.
 
Discussions of lensing effects of different objects can also be found
\cite{SL00,KT96}.

\subsection{Gravitational microlensing and MACHOS \label{macho}}
 
A statistical analysis of the {\em galactic halo} via microlensing
\cite{Pa86,Su99} suggests that MACHOs (massive compact halo objects)
account for a significant part ($>$ 20\%)
of the total halo mass of our galaxy. Their most likely mass range
seems to be in the range
0.3 -- 0.8 $M_\odot$, with an average mass of 0.5 $M_\odot$,
cf.~\cite{J96,Su99}.
If the bulge is more massive than the standard
halo model assumes, the average MACHO mass \cite{J96}
will be somewhat lower at $\sim 0.1$ $M_\odot$.
However, there are some astrophysical difficulties with this
result of estimated $\sim 0.5$ $M_\odot$ for the lenses.
These cannot be hydrogen-burning stars in the halo since
such objects are limited to less than 3\%\ of the halo mass by
deep star counts \cite{GBF97}. Modifying the halo model to slow down
the lens velocities can reduce the implied lens mass somewhat, but
probably
not below the sub-stellar limit 0.08 $M_\odot$.
Old white dwarfs have about the right mass and
can evade the direct-detection constraints, but
it is difficult to form them with high efficiency, and there may be
problems with overproduction of metals and
overproduction of light at high redshifts from the luminous
stars which were the progenitors of the white dwarfs \cite{CS95}.
Primordial BHs are a viable possibility, though the coincidence
has to be explained to have them in a stellar mass range.
 
\begin{figure}[th]
\centering
\leavevmode\epsfysize=7cm \epsfbox{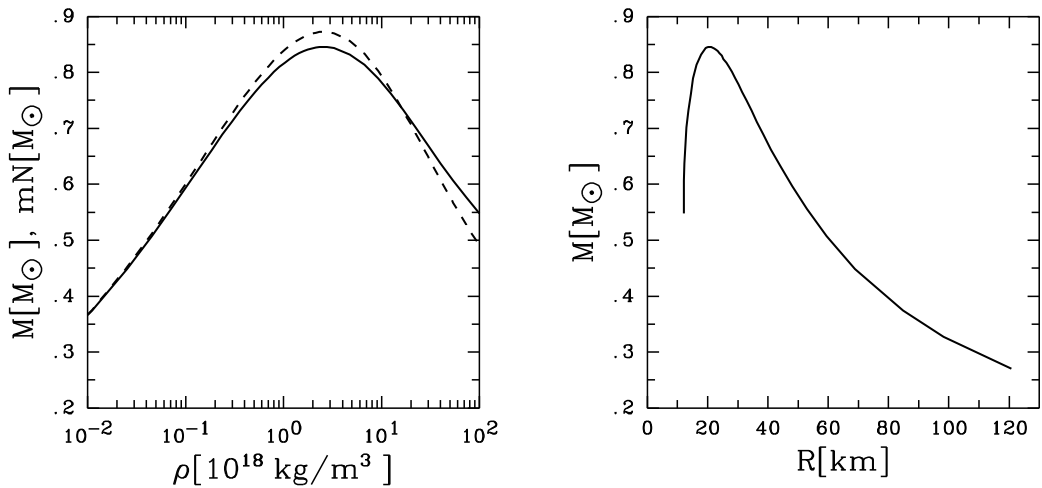}\\
\caption[fig2]{Left: Mass $M$ and particle number $N$ (or
rest mass $mN$ at infinity) of a mini-BS depending on the central density
$\rho$. Right: Mass-radius dependence of a mini-BS.\label{fig2}}
\end{figure}

Due to these difficulties of getting MACHOs in the inferred mass range
without violating other constraints, there was the suggestion
that BSs could be the explanation \cite{MS00}.
The explicit physical values for a mini-BS
has been exhibited in Fig.~\ref{fig2} with a
scalar field mass of $10^{-10}$ eV/$c^2$.

At the left-hand side of Fig.~\ref{fig2}, the dependence of
the mass $M$ and the particle number $N$ (or rest mass)
on the central density $\rho$ is shown.
Stable non-rotating BSs exist at the lower
central densities below the maximum mass.
The critical values are $M_{\rm crit}=0.846$ $M_\odot$ and
$mN_{\rm crit}=0.873$ $M_\odot$ for a central density of
$\rho_{\rm c}=9.1\times \rho_{\rm nucl}$,
where $\rho_{\rm nucl}=2.8\times 10^{17}$ kg/m$^3$ is the average
density of nuclei. Since non-interacting bosons are very ``soft"
(due to Bose-Einstein condensation), BSs are extremely dense objects
with a critical density higher than comparable ones of neutron or
strange stars \cite{GKW95}. The figure on the right-hand side gives the
mass depending on the radius.
For the mass-radius diagram,  99.9\%\
of the total mass was chosen as effective radius. This ensures that
the exponentially
decreasing exosphere of the BS has almost no influence on the
asymptotic Schwarzschild spacetime.
 
Stable BSs have radii larger than the minimum at 20.5 km
with a mass of 0.846 $M_\odot$. The choice for the boson mass $m$
yielded that the total mass of these relativistic BSs is
just in the observed range of 0.3 to 0.8 $M_\odot$
for MACHOs.
 
As was the case in the discussion of axidilaton stars (Section
\ref{axisect}),
one could  turn this argument around: By identifying
the most massive MACHOs with known gravitational mass
of about 0.8 $M_\odot$ as BS, one is essentially `weighing', via
$M_{\rm Kaup}/N_{\rm crit} \cong m$, the boson mass to
$m \sim 10^{-10}$ eV/$c^2$.

\subsection{{\v C}erenkov radiation from boson stars}
 
As in Section \ref{lensing}, we assume that the BS interior is empty,
but now electrically charged particles move fast through the
gravitational
BS field. In general, if a particle moves through a medium with
a constant velocity greater than the velocity of light in that medium
then {\v C}erenkov radiation occurs. Because of the superluminal
particle motion, a shock wave is created and this yields a loss
of energy.
 
Accelerated charged particles emit electromagnetic radiation, but
acceleration only by a gravitational field does not produce
{\v C}erenkov radiation, in agreement with the equivalence principle.
There, the acceleration induced by gravity disappears in the local
inertial frame (further details in the references of \cite{CLT00}).
Despite that, the gravitational BS field can act as a medium with
an effective refractive index $n_\gamma$ for light
if particles move with constant velocity through it \cite{GMS99}.
It is found
\be
n_\gamma^2 (k_0) =
  1 - \frac{1}{k_0^2}R^i_{\ i} \; ,
\ee
where $k_0$ is the frequency of the emitted photon $\gamma$ and
$R^i_{\ i}$ is the sum on the spatial indices of the Ricci tensor.
{\v C}erenkov radiation is now kinematically allowed if
$R^i_{\ i}<0$, i.e.~$n_\gamma^2 (k_0)>1$.
 
{}Following \cite{GMS99}, it has been  investigated in \cite{CLT00}
whether  BS models allow {\v C}erenkov radiation.
It was found that stable mini-BSs and stable BSs cannot
generate {\v C}erenkov radiation, whereas unstable BSs can.
However, non-topological soliton stars do allow this effect
where the position and the strength depend on the central
scalar field value and the potential parameter $\Phi_0$.
Then, {\v C}erenkov radiation appears in spherical shells
and only there, but this can be everywhere inside the star.
 
Again, we recognise how important the influence of the scalar
field potential on detection can be.

%************************
\subsection{Gravitational waves \label{waves}}
 
In the early stage of BS formation,
highly excited configurations are expected to exist in which
the quantum numbers $n$, $l$ and $a$ of the gravitational atom,
i.e.~the number $n-1$ of nodes, the angular momentum and the
azimuthal angular dependence $e^{-i a \varphi}$ are non-zero;
cf.~\cite{Je89}.
For energetic reasons, these excited BSs are presumably only meta-stable,
all initially higher modes during the BS formation
have eventually to decay into the ground state $n=1, l=0$
(with $a=0$ for non-rotating and $a\ne 0$ for rotating BSs)
by a combined emission of scalar matter radiation
and gravitational radiation.
 
If we assume a BS to be in an excited state with $l=a=0$, we know that
its mass and particle number depend linearly on the number of nodes
$n-1$, cf.~\cite{FLP87a,BG87}. Since these states have zero quadrupole
moment,
the only decay channel is scalar matter radiation.
For the transition of an excited BS with only $n>1$
into the ground state, it is reckoned \cite{FG89} that the energy
released by
scalar radiation is about
\be
E_{\rm rad} \sim (n-1) M_{\rm Pl}^2/m\; , \qquad
\Delta N\sim (n-1)  (M_{\rm Pl}/m)^2\, ,
\ee
where the loss of bosonic particles is given by $\Delta N$.
 
Ferrell and Gleiser \cite{FG89} considered the
situation of a Newtonian BS that has the main part of particles in
the $1s$ state and only a small number in the $3d$
($0.286$ parts per thousand). They estimated the
amount of total energy released by the transition from this small number
of particles in the $3d$ into the $1s$ state.
The lowest mode which has quadrupole moment and therefore
can radiate {\em gravitational waves} is the $3d$ state
with $n=3$ and $l=2$. For the $\Delta n=2$, $\Delta l=2$ transition,
the BS will decay into the $1s$ ground state with $n=1$ and $l=0$, and
thereby the gravitational radiation process preserves the particle
number $N$. Even though
only a relative small number of particles populates the $3d$ state,
the radiated energy is quite large \cite{FG89},
i.e., $E_{\rm rad} =2.9 \times 10^{22}$ (GeV/$mc^2$) Ws.
The typical frequency of this gravitational radiation is
$\nu_{\rm g}=2.03 \times 10^{22}$ ($mc^2$/GeV) Hz with the power of
$1.3 \times 10^{39}$ W
but much less than the output of a supernova explosion of
about $10^{43}$ W. Highly relativistic BSs with radius near
the Schwarzschild one ($M/R\sim 1/G$) could have a maximal power of
$10^{52}$ W, but this is just a theoretical upper bound on the
gravitational radiation of a self-interacting system.
In the scenario of \cite{FG89}, it is concluded that the
excited state will decay extremely fast to the ground state with the
exponential decay time of about $1.1 \times 10^{-16}$ ($mc^2$/GeV) s.
Thus, the final phase of the BS formation would terminate in an
outburst of gravitational radiation despite the smallness of the object,
cf.~Table IV.

In \cite{DD03}, the collapse scenario of parts of a
charged boson cloud with $\omega=m$, i.e., just above the
eigenfrequencies of excited CBSs, is investigated.
In comparison with the results above,
gravitational waves are created in a shorter
time interval ($10^{-25}$ s) and with much less
power ($4 \times 10^{14}$ W).
 
The issue of non-radial pulsations of a BS was mathematically
formulated \cite{KYF91}, whereas quasinormal modes are explicitly
calculated in \cite{YEF94}.
The interest in non-radial pulsations arises from the fact
that they transport gravitational waves away from the BS.
{}For linearised perturbations around an equilibrium BS,
odd and even parity modes can be distinguished.
It was shown that the odd parity modes do not couple to gravitational
waves as well as for the stellar pulsation of a perfect fluid.
This mode type manifests the propagation of the gravitational
wave through the star, it does not change the star's density or pressure
distributions. With a different analytical approach, the even-parity
quadrupole oscillations ($l=2$ modes) which do couple to gravitational
waves were investigated numerically in \cite{YEF94}.
These quasinormal modes describe
oscillations with emitted gravitational and scalar matter radiation
(but no incoming radiation). In contradistinction to
relativistic fluid stars, the imaginary parts of the frequencies are
large, i.e.~the $f$ and $p$ modes imply very short damping time scales.
It is also found that the amplitude of the metric perturbation is
comparable to the one of the scalar field, again different to the
$w$ modes of an ordinary star.
No weakly damped modes are present within the BS. One restriction of
the numerical calculations in \cite{YEF94} is connected with the
definition of the BS surface. Due to numerical reasons, a clear surface
had to be defined, but a BS has only an exponentially decreasing
exosphere;
this has a crucial influence on the values of the eigenfrequencies;
an introduction to pulsating relativistic stars can be found in
\cite{Ko96}.
 
Future gravitational-wave detectors may have also different ways of
searching for BSs.
If a compact object of, say, a solar mass is observed to be spiralling
into a central one with a much larger mass, one would be able to test
the larger object. From the emitted gravitational waves, the values
of the lowest few multipole moments of the central object
can be extracted, like mass $M$, angular momentum $J$,
mass quadrupole moment $M_2$.
For a BH as central object,
the no-hair theorem states that all its multipole moments are uniquely
determined by the two lowest ones, e.g.~$M_2=-J^2/M$. In \cite{R97}
$M,J,M_2$ and the spin octopole moment $S_3$ are calculated for
rotating BSs in distinction to a BH.
For example, if one observes the combinations
$J/M^2=0.01$, $-M_2M/J^2=24$, and $S_3M^2/J^3=19$, then
the massive central object would be very likely a rotating BS.
If instead, one finds $-M_2M/J^2=1$ and $S_3M^2/J^3=1$,
then a BH is the likely source, and if one determines
$-M_2M/J^2=24$ and $S_3M^2/J^3=4$, a different configuration
(BS with different kind of self-interaction, e.g.) may be detected.
Further numerical combinations for BSs result from the Figures in
\cite{R97}.
 
For a rotating BS, co-rotating particles orbiting with circular
geodesics in the equatorial plane were investigated in \cite{R97}.
The particle energy depending on its gravitational-wave
frequency for one BS solution is shown there in Figure 6. If particles
could travel freely inside the BS, it would, in principle, be possible
to map out the interior of the BS via the emitted gravitational waves.
 
If a BS experiences some kind of perturbation, we may be able
to measure characteristic frequencies.
In \cite{HC00}, the mode frequencies for the fundamental mode and the
first harmonic mode of radial perturbations of mini-BSs are exhibited
in Tables II and III (cf.~Figs.~7 and 8 as well); cf.~Fig.~6 in
\cite{SeSu90} where similar frequencies were discovered.
Further details of Ref.~\cite{HC00} are mentioned in Section
\ref{catastrophe}.
 
The output of gravitational waves
was considered in \cite{Gl89} for identical BS binary systems,
assuming that they are located in
galactic halos. However, the gravitational radiation background
contains data from ordinary binary systems as well.
The estimation for critical BSs gives a
contribution to the gravitational radiation background where
the amplitude is $h \sim 3\times 10^{-25}$ (TeV/$mc^2)^{3/2}$
and the frequency $\nu \sim 25$ (TeV/$mc^2)^{5/4}$ Hz.
This result is valid if the BS binary density is about the critical
density and they dominate over ordinary binaries if $m \le 10^6$
TeV/$c^2$.
For mini-BS binaries with scalar field mass $m=10^{-5}$ eV/$c^2$,
the typical amplitude is $h \sim 10^{-24}$
and the frequency $\nu \sim 4.5$ Hz, again for a mini-BS binary density
of about the critical density.

\subsection{Boson star in the galactic centre? \label{galaxy}}
 
Several  galaxy centres contain $10^6$ up to 10$^9$ solar masses.
The widely accepted explanation is
a supermassive BH or a rich cluster of objects (stars, BHs, e.g.).
Already in \cite{SL97,SL98},  a so-called mini-BS or BS has been proposed
as an alternative explanation, cf.~Sections \ref{curves}
and \ref{redshift}.
 
In \cite{TCL00}, a supermassive mini-BS as model for the Galactic
centre was compared with the observational data available for our Galaxy
where,
at least, the existence of a star cluster can be excluded
\cite{EG96,EG97}; cf.~\cite{GKMB98}.
The central object has a mass of
about $2.6 \times 10^6 M_\odot$. The more recent observations
of \cite{GKMB98} probe the gravitational potential at a radius
larger than $4 \times 10^4$ Schwarzschild radii of such a BH.
The mentioned data consist of the movement of
stars around the Galactic centre and, in \cite{TCL00},
they were explicitly fitted by a solution of a
mini-BS with $m \sim 10^{-17}$ eV/$c^2$ plus a stellar cluster.
The same result can be received by using a BS solution with
$m \sim 10^{-4}$ GeV/$c^2$ ($\lambda=1$) and a non-topological
soliton star solution ($\Phi_0 \sim m$) with $m \sim 10^{4}$ GeV/$c^2$.
The data can be fitted by a central BH as well and, so far,
no distinction can be obtained between
a BH and a BS at the centre. The difference starts at a radius more than
three orders of magnitude below the innermost data point, in case of
a general relativistic mini-BS
\cite{TCL00}. A mini-BS in the Newtonian approximation has less mass,
but if we reduce the constituent mass $m$ in order to obtain
the same central mass,
the radius becomes about two orders of magnitude larger (cf.~Fig.~5
and Table 2 in \cite{SL97}).
Observations are just one to two order of magnitudes away from
finding differences in the BH-BS comparison.
Let us also mention that  an extended neutrino ball can
explain the data \cite{TV98}  so far.
 
A non-rotating BS may have also the problem of accreting too much matter
and so forming a BH eventually.
It is obvious that inspiralling interstellar gas and stars will
collide with each other, glue together within the BS, and so
eventually form a BH. If instead a small BH spirals inwards and stays
there, the final state of a non-rotating BS will be without doubts a BH.
The investigation of the effective potentials for massless and massive
particles of a spherically symmetric mini-BS can give some understanding
how infalling matter behaves inside the BS \cite{TCL00}.
All particles with orbital angular momentum and unbound orbit do not
move through the BS centre, hence, leave the star; only if the orbital
angular momentum is zero the particles will move exactly through the
centre.
In this ideal situation of non-interacting matter,
every matter is removed from the BS interior
and no BH can form. But, as mentioned above, presumably
matter will collide and glue together so that a BH cannot be avoided
(i.e., the particles have bound orbits).
However, in case of a rotating BS, there is the possibility that matter
could be guided along geodesics on the rotation axis forming jets;
this has still to be shown. {}For a
BS in the limit $\Lambda \gg 1$, circular orbits are stable until
about 5/3 times the radius of the doughnut hole \cite{R97}.
 
On the basis of an effective potential, it is shown \cite{T02} that,
for a spherically symmetric mini-BS,
circular orbits exist for every value of the radial coordinate and are
stable, even inside the BS. Hence, accretion could follow a series of
stable orbits and finally end up at the centre. Further
investigations are needed in order to clarify whether
a BH forms.
 
What happens if the BS under external perturbations (gravitating scalar
matter) is forced to leave the equilibrium ground state
changing mass and particle number in the configuration?
The numerical simulations of \cite{SeSu90,SeSu94,BSS98,B99} show that
stable mini-BSs under finite non-infinitesimal perturbations
start to oscillate, emit scalar field radiation, and settle down
into a new mini-BS configuration with less mass and a larger radius.
If the amount of scalar accretion matter does not exceed a critical value
during some time, the BS is not jeopardised to form a BH.
 
The emissivity properties of a geometrically thin, optically thick,
steady
accretion disc with constant accretion rate around a supermassive
critical mini-BS are demonstrated in
\cite{T02} and compared with a BH of same mass. Optical thickness means
that the local effective temperature must be sufficient to
radiate away the local energy production.
The disc calculations are only valid where the gravitational potential is
$\sim GM/r$, i.e.~outside of the investigated mini-BS.
One result is that, for the inner disc parts,
a BS produces more power per unit area and a hotter disc than the
comparable BH; for this critical mini-BS, there is a temperature
difference of
about two thousand degrees in the inner parts of the accretion disc.
For a BS with $\Lambda=100$, there is about the same maximal
temperature but the position of that maximum is closer to the BS centre.
The calculation of the final emission spectrum reveals a difference
between the mini-BS and the BH model in the far ultraviolet regime
$>10^{16}$ Hz (Fig.~3 in \cite{T02});
this disc model shows no production of X-ray or gamma-ray energy.
The Eddington luminosity, at the balance of
the inward force of gravity with the outward pressure
of radiation, is not exceeded within the mini-BS interior. Thus,
accretion could continue inwards a mini-BS.
A further result \cite{T02} is that the disc model does not describe the
Galactic centre Sgr A$^\star$ neither for a BH nor a BS due to the
accretion disc properties itself. It produces several order of
magnitudes too much accretion luminosity. In contrast to observations
is also the peak of the standard accretion disc broad band
spectrum in the infrared.
 
Additionally, BSs could disrupt stars \cite{TCL00}. In general,
there is a distance to a central object where the
extension of a normal star cannot be neglected and where
tidal forces on the star become important. The corresponding
so-called tidal radius is defined where $M/r^3$ is equal to the mean
internal energy of the passing star. Estimations show that
BHs should have masses below $10^8 M_\odot$ so that they provoke
disruption outside the event horizon. In general, such an event takes
place
once in about $10^4$ yrs. Hence, whereas a too massive BH disrupts stars
just inside its event horizon, a BS produces this effect visible
for observations (inside or outside of its effective radius).
 
Observations in the x-ray band within the central regions of active
galaxies
have shown an interesting feature, a Doppler and gravitationally
redshifted iron K$\alpha$ line \cite{Ta95,FN95}. It was claimed
\cite{Iw96} that the data can be understood as being emitted from within
1.5 Schwarzschild radii near a rotating BH. In \cite{SL97,SL98}, we
speculated that a BS geometry may be able to explain the data as well.
Recently in \cite{LT02}, a detailed numerical analysis of a
geometrically thin, optically thick accretion disc around a
supermassive critical mini-BS simulated this situation for different
inclination angles. It is concluded that the observed line profile
of the galaxy MCG-6-30-15 cannot be fitted by a spherically symmetric
mini-BS or BH for the chosen accretion disc model.
For a possible identification of such a non-rotating mini-BS,
the characteristic spectral feature
would be the detection of an intense double peak for face-on galaxies
or, for larger inclination angles, of several peaks
of the long drawn-out K$\alpha$ line.

%***************
\subsection{Boson star inside a H{\footnotesize II} cloud?}
 
BSs could exist in two more scenarios as it was pointed out in
\cite{SL97,SL98}:
 
Within
host galaxies of quasars, bright H{\footnotesize II} regions are observed
\cite{BKS96}. By imagining that neutral H{\footnotesize I} gas clouds
could gain kinetic energy in the gravitational potentials of BSs,
an excited hydrogen gas cloud, i.e.~H{\footnotesize II}, results.
 
The second situation is that a light BS is completely contained within
a nuclear burning star. In this case, both stellar structures
would be interacting mutually, especially changing the model parameters
of the conventional star. This resembles the so-called Thorne-\.{Z}ytkow
object, i.e., a neutron star core inside a supergiant \cite{TZ77}.
If instead the BS matter has some kind of interaction with high
energetic photons, it might be possible that the BS dissociates
and the remaining scalar particles circulate freely within the
conventional star.
A so-called {\em cosmion} represents such a scenario and tries to explain
the solar neutrino problem \cite{FG85}.

%*************************************************
\subsection{Gravitational memory and the evolution of boson stars}
 
The evolution of conventional nuclear burning stars follows
the Hertzsprung-Russell diagram of the luminosity  as a
function of the temperature. During its life time, such a star
has different stages which are expressed by a line in this  diagram.
There is the possibility that a BS could go through a continually
evolution as well, if the  gravitational attraction changes during the
evolution of the Universe.
 
In JBD or scalar tensor theory, a real scalar field regulates the
strength
of the gravitational force, cf.~Section \ref{JBD}.
Suppose that at the time of
BS formation, the gravitational strength was different from the
one prevailing today; following this theory, $G$ was larger in the past.
The real scalar field is actually a radial function
so that the gravitational strength {\em inside}
the JBD-BS is determined by a mutual
gravitational interaction of the complex and the real scalar field
matter.
These investigations  are outlined in
Section \ref{JBD}. From the observational side, it may be important that,
inside the BS,
still the strength of the gravitational constant at formation time
could be effective \cite{TLS98}. According to this
{\em gravitational memory effect}, the BS is book-keeping the evolution
of $G$.
Because the JBD field is a radial function and changes its value
at infinity, there should be a repercussion on the BS. But if this
change is much slower than the cosmological evolution of $G$,
the star is practically static. Due to this effect, physical
properties like the radius depend on formation time and could be
different even for the same total mass.
 
If instead the JBD field adapts quickly inside the JBD-BS as changes
occur
at infinity, the BS evolves as well.
Figure 1 in \cite{TSL98} shows the mass against the value
of the real scalar field at infinity for constant particle number.
It is derived that the JBD-BS mass increases with increasing
$\phi_{\rm JBD}(\infty)=1/G$, i.e.~with time.
One can understand that
an increasing JBD field pumps energy into the BS and
increases the mass. Thereby, the {\em evolution} of a BS in
JBD theory demonstrated in this Figure 1 resembles the evolution
in the Hertzsprung-Russell diagram of conventional stars.
The physical properties as mass, radius, and stability is
influenced by this repercussion \cite{TSL98}, such that,
for example, the mass changes about 2 \%.

%******************************
\section{Real scalar field boson stars \label{RSFBS}}
 
In contrast to the complex case, real scalar fields  do not
possess a conserved charge or particle number. Moreover,
a static configuration $\Phi (r) = P(r)$, i.e.~(\ref{statio}) for
$\omega=0$,
would, in flat spacetime, be unstable due to Derrick's theorem
\cite{De64}, whereas
a superposition of positive and negative frequency states
$\phi_1 (r,t) = \left(P(r) e^{-i\omega t}
   + P(r) e^{i\omega t}\right)/\sqrt{2}$$=\sqrt{2} P(r) \cos(\omega t)$
 is time-dependent and implies time-dependence of the
energy-momentum tensor and the
metric, as well. These configurations are
called {\em oscillating soliton stars} \cite{SeSu91,SeSu94} or
`oscillatons'
\cite{Ur02,UMB02} and are, in general, unstable. Actually, our global
stability analysis via catastrophe theory \cite{KMS91,KMS91b}
has revealed already earlier the existence of oscillating BSs,
as we will call them in the context of this review.
Q-stars are a class of non-topological solitons
which are already discussed in detail in \cite{Je92};
cf.~\cite{Co85,Ly89,SKL89,BLS90,SSP92}. Further interesting
configurations are the so-called {\em pulsons} \cite{BM76,BM77} or {\em
oscillons}
\cite{HoC02}, respectively; these are localised, time-dependent,
unstable, spherically symmetric solutions to a nonlinear KG equation.
 
The temperature $T$ is the decisive parameter characterising the
thermodynamic equilibrium of such a boson gas of real scalar particles
\cite{Li79}. The particle density in momentum space is given by
\be
n_{\Phi} = \frac{1}{\exp (E_{\Phi}/k_B T) - 1} \; ,
\ee
where $E_{\Phi}=\sqrt{{\bf p}^2+m^2}$ is the energy of the particle
with momentum ${\bf p}$ and mass $m$. In the limit of  vanishing
temperature,
one obtains $n_{\Phi}=0$. Hence, a BS consisting of real scalars
cannot exist for $T=0$ and a different method
has to ensure a conserved particle number.
 
{}For a massless real scalar \cite{Ch86},
analytical solutions of the Einstein-KG system are derived
first by Buchdahl \cite{Bu59}, cf.~Wyman \cite{Wy81} and
Baekler et al.~\cite{BMHH87}.
Wyman showed that all static spherically symmetric
solutions are contained in the two-parameter family of Buchdahl.
The common characteristic of all such
solutions is that they are singular or topological non-trivial
\cite{Ko78,KOS79}.
In \cite{ScSc91,Sc91}, the ADM mass of such solutions were calculated
because the metric is asymptotically flat and the scalar field vanishes
at infinity. Since the latter has a logarithmic singularity at the
centre,
for some solutions, the metric is singular as well and these
are typical examples for {\em naked singularities}.

By including a mass term and a quartic self-interaction,
numerical solutions of the Einstein-KG system can be determined
\cite{JS92}. Again the scalar field possesses a singularity at the
origin;
alternatively, one finds regular solutions at the centre, but
the singularity is shifted to a finite distance. The
dynamical stability investigation used a variational perturbation method
estimating upper bounds for an eigenvalue. The result showed
that all such naked singularities are unstable \cite{JS92,JS97}.
However, a direct numerical determination of the ground state
eigenvalue indicates perturbatively stable solutions \cite{CDL98}.

In the numerical one-parameter-solutions of Ref.~\cite{JS92},
the `bare' mass $m$ is the decisive parameter, whereas $\lambda$
is not important. In the case of complex scalars, two parameters occur:
besides $m$, we have the frequency $\omega$, and globally regular
solutions exist. Thus,
the parameter $\omega$ in the phase factor is actually responsible for
the regularity of the solutions. This is not surprising because this
parameter is proportional to the conserved particle number $N$.
This consideration motivated the construction of
the globally regular one-parameter-solutions with non-vanishing
$\omega$ and $m=0$ in \cite{Sch94,Sch95,Sch97,Sch98,Sch98b};
cf.~boson halos in Section \ref{halo}.

%************************************
\subsection{Quantised real scalar fields \label{Q-RSF-BS}}
 
An assembly of real scalar particles experiencing just a mass term
was already considered in 1966 by Bonazzola and Pacini \cite{BP66}.
In this approach, the scalar field is second quantised.
Later on, solutions for this model of ``systems of
self-gravitating'' bosons were considered by Ruffini
and Bonazzola \cite{RB69} similarly to the Hartree-Fock atom.
Because a real scalar has no {\em anti-particle} states,
the corresponding KG field in a spherically symmetric spacetime metric
can be merely  decomposed into positive and negative
frequency field operators
\be
\Phi=\Phi^+ + \Phi^- \, ,
\ee
where
\ba
\Phi^+ & = & \sum_{nla} b_{nla}^\dagger R_l^n(r)
 Y^{|a|}_l(\theta, \, \varphi)
  e^{-i \omega_{nl} t}\,,\nonumber \\
  \Phi^- & = & \sum_{nla} b_{nla} R_l^n(r)
 [Y^{|a|}_l(\theta, \, \varphi)]^*
  e^{+i \omega_{nl} t}
\ea
are  again generalisations of the wave function
of the hydrogen atom for a {\em bound  state}.
Here $R_l^n(r)$ are radial distributions,
$Y^{|a|}_l(\theta, \, \varphi)=(1/\sqrt{4\pi})
P^{|a|}_l(\cos\theta)\, e^{-i|a|\varphi}$ the spherical harmonics
given in terms of the
normalised Legendre polynomials, and $|a|\leq l$ are  the quantum numbers
of {\em azimuthal}  and {\em angular} momentum. (The nodeless radial
field
solution $R_{0}^0(r)$ is earlier denoted by $P(r)$.)
The non-vanishing {\em b}osonic commutation relations are
\be
[b_{nla},b_{n'l'a'}^\dagger] = \delta_{nn'} \delta_{ll'} \delta_{aa'} \,
.
\ee
For the ground state of a cold configuration,
\be
|N\rangle=|N,0,0,\ldots\rangle:=\prod_1^N b_{100}^\dagger \vert0\rangle
\ee
is chosen \cite{RB69},
where $\vert0\rangle$ is a vacuum state in the curved spacetime
`background'.
The BS energy-momentum tensor $T_{\mu \nu}(\Phi)$ becomes now an
operator.
Hence, for the right-hand side of the Einstein equation
(\ref{phi152}), the vacuum expectation value
$\langle T_{\mu \nu}\rangle:= \langle N|:T_{\mu \nu}:|N\rangle$
is calculated for the ground state where
$:T_{\mu \nu}:$ denotes normal ordering of the operator products.
 
An {\em excited state} of positive energy  can be defined by
$\vert N,n,l,a\rangle:= \Phi^+ \vert0\rangle$.
Such a `gravitational atom' \cite{FG89} represents a {\em coherent}
quantum state, which nevertheless can have macroscopic size and large
mass.
The gravitational field is self-generated
via the mean value of the energy-momentum tensor, but remains
completely classical, whereas the real scalar is treated to some
extent as operator.
 
As we noted in the beginning of this Section,
there is no conserved particle number for real scalars, and so one needs
to
introduce the normalisation condition
\be
2\pi  \int \limits_0^\infty  r^2 e^{(\lambda - \nu )/2} \;
  \bigl [ \omega_{nl} R_{nl}(r) + \omega_{n'l'} R_{n'l'}(r)\bigr ] \; dr
=
\delta_{nn'} \delta_{ll'}
\ee
in order to stabilise the system. Similarly as for complex scalars,
the constraint is imposed that
the solutions are asymptotically flat at spatial infinity.
 
Due to the fact that the mean value of the energy-momentum tensor is
employed, the same system of differential equations
(\ref{phi152}), (\ref{phi153}) arises as for the complex case.
A later investigation \cite{BGZ84}  rediscovered these results.
 
An interesting issue is how a BS reacts to a
second-order phase transition in the early Universe.
In \cite{HKL02}, such a spontaneous symmetry breaking
has been investigated by assuming a negative quadratic term in
the scalar potential.
After the phase transition, the `bare' mass $m$ is increased
and the self-interaction strength is reduced by
$\Lambda \rightarrow \Lambda/2$, resulting
in a smaller total mass $M$ of the BS and,
thus, preserving its existence at all.
 
In \cite{HLM89,HLM90,LH92}, the BS of a field quantised
real scalar, the model of Ruffini and Bonazzola \cite{RB69}, has been
extended by including a neutron star in perfect fluid description.
The details are discussed in the 1992 reviews.
 
The situation of a field quantised complex scalar is exhibited
in Section \ref{Q-CSF-BS}.

%****************************************
\subsection{Dilaton star}
 
The most general scale-invariant Lagrangian density for the real
dilaton field $\varphi$ coupled to gravity adopts the form
\be
{\cal L}_{\rm dil} =  {\sqrt{\mid g\mid }}
 e^{2 \varphi /f_\varphi} \left \{\frac{1}{2\kappa } R + \frac{1}{2}
   g^{\mu \nu} (\partial_\mu \varphi ) (\partial_\nu \varphi )
    - \frac{\lambda}{4} f_\varphi^4 e^{2 \varphi /f_\varphi}
 \right \} \, ,
 \label{lagrGK}
\ee
where $\lambda $ is a coupling constant and $f_\varphi$ is the
'decay constant' of the dilaton. The dilaton mass itself is given by
$m_{\rm dil}=2\sqrt{\lambda} f_\varphi$. For this non-minimal coupling,
{\em scale invariance} is realized by
the conformal change  $g_{\mu \nu} \rightarrow \exp(-2\alpha) g_{\mu
\nu}$
and the shift $\varphi \rightarrow \varphi + \alpha f_\varphi$ in the
dilaton field,
where $\alpha$ is some dimensionless parameter.
If $f_\varphi=\sqrt{3/8\pi} m_{\rm Pl}$, the theory is invariant under
the
larger symmetry group of conformal transformations, induced by a
space-time dependent parameter $\alpha= \alpha(x)$.
For a scale-invariant interaction,
the potential
$V(\varphi)= e^{4 \varphi /f_\varphi} \lambda f_\varphi^4/4$ is the only
allowed type. For any other dilaton potential,
the theory is still of JBD type in terms of $\phi=
f_\varphi e^{ \varphi /f_\varphi}$, but not anymore scale invariant.
In the Einstein frame, the theory consists just of a massless
real scalar plus a cosmological constant
$\Lambda = \lambda f_\varphi^4/4$ coupled to gravity.
 
Since the Lagrangian is scale invariant, Noether's theorem provides the
conserved dilaton charge
\be
Q_{\rm dil} = \left(1+6/\kappa f_\varphi^2\right)
  \int \sqrt{\mid g\mid } \, e^{2 \varphi /f_\varphi}
  \partial^t \varphi \;  d^3 x  \, .
\ee
It vanishes if $\varphi$ is time-independent as in the case of
a static space-time of a non-rotating star. Therefore, in order to
construct
static {\em dilaton stars}, Gradwohl and K\"albermann \cite{GK89}
followed the semi-classical approach of Ruffini and
Bonazzola \cite{RB69} and Breit et al.~\cite{BGZ84} for real scalars.
They regularise the energy-momentum operator $P_\mu$ by means
of the subtraction $P_\mu':= P_\mu- \langle 0 \vert P_\mu \vert 0
\rangle$
of its vacuum expectation value. In curved spacetime,
this Wick or normal ordering of the
operator products, should be performed by the point-splitting method,
in order to avoid ambiguities due to the non-uniqueness of the
vacuum state, cf.~\cite{Wa94}, p.~87.
Thereby, they can effectively get rid of the cosmological constant.
A spontaneous breaking of the
scale invariance leads to a vanishing dilaton expectation value
at infinity. As a consequence, GR is recovered asymptotically
and the {\em dilaton star} carries zero dilaton charge,
but has a constant particle number $N$ due to the normalisation
of the dilaton field.
 
The following parameter combinations have been considered:
$\overline\lambda :=\lambda(f_\varphi/\omega)^2= 1, 10, 30, 100$ and
$\alpha=(f_\varphi /M_{\rm Pl})^2 = 1/40, 3/(8\pi), 1, 10$. The most
noticeable effect on the total mass $M$ is due to the change
in the latter parameter.
For a very  light dilaton of mass $m_{\rm dil}=10^{-11}$ eV/$c^2$,
Gradwohl and K\"albermann \cite{GK89} found as maximal values
\be
M_{\rm crit} = 7\sqrt{\lambda\;} \frac{f_\varphi}{\omega} M_\odot\, ,
\qquad
R_{\rm crit} = 40\sqrt{\lambda\;}\frac{f_\varphi}{\omega} \; {\rm km}\, ,
\ee
where $\omega$ is the ground state frequency.

%**********************************
\subsection{Axion star}
 
In general, axions have no effective cooling mechanism;
they remain in the form of an extended gas cloud after separation
from the Hubble flow. Decay processes may help in cooling down an
axion cloud. In a cosmological context, there is a temperature dependence
of the axion decay modes\cite{PWW83}.
The stimulated decay process $aa \rightarrow a \gamma\gamma$,
together with $aa \rightarrow \gamma\gamma$ and $aa \rightarrow \nu\bar
\nu$ have been examined for their importance in the cooling of an
axion cluster \cite{FTY85}.
(The on-shell contribution $aa \rightarrow aa$ followed by
$a \rightarrow \gamma\gamma$ can be ignored because this does not yield
an
energy loss.) As later shown in \cite{T86}, the stimulated
processes can only help in forming axion stars,
by including the effect of large phase-space density;
this process relaxes the axion clouds efficiently to form
axion stars. On the other hand, free axions have a lifetime which is
longer than the age of the universe so that the axion stars can be
re-powered by accretion. Axion stars could be
detectable as coherent cosmic masers \cite{T86}. In the same publication,
an axion star for a real scalar field $a$ with potential
$m^2a^2+\lambda a^4$ is investigated resulting in a critical
mass in the order of magnitude of $M_\odot$. Hence,
for a real scalar field, we obtain the same result
as  Colpi et al.~\cite{CSW86} in the case of complex scalars.
For this potential,
the axion star may radiate as a radio source. Additionally, the
relaxation time of a gravitationally bounded cloud of axions as
well as the luminosity of the axion star is estimated in \cite{T91}.
 
The formation of axion stars at the QCD epoch was
considered in \cite{HR88}. The initial isocurvature perturbations
yield small axion miniclusters of mass about $10^{24}$ kg with a
diameter of the order of $10^{12}$ m. In later evolution, the
miniclusters undergo standard hierarchical gravitational clustering.
For the minicluster density of about $10^{10}$ kg/m$^3$,
axion annihilation $aa\rightarrow \gamma \gamma $
is not important, but in axion stars with densities of about
$10^{27}$ kg/m$^3$, it makes the configuration unstable
\cite{SeSu94,DW96}. More recently, the formation of axion halos
is discussed in \cite{HBG00}; cf.~the constraints on scalar parameters
in \cite{RT00}.
 
Numerical studies \cite{KT93} including nonlinear effects in the
evolution of inhomogeneities in the axion field around the QCD epoch
can lead to very dense axion clusters with densities high enough so
that the stimulated decay process supports to lead to Bose-Einstein
condensation, and eventually, to axion stars.
Femto- and picolensing by axion miniclusters is investigated
in \cite{KT96}.
 
A different way how axion stars may send signals are exhibited
in \cite{Iw98,Iw99a,Iw99b,Iw99c,Iw99d,Iw99e,Iw00}.
There, the {\em oscillating soliton}
star model \cite{SeSu91} is taken as example for axion stars. Since
axions couple to the electromagnetic field,
axion stars could dissipate their energies in magnetised
conducting media surrounding white dwarfs or neutron stars.
Oscillating configurations may generate monochromatic
radiation with an energy equal to the axion mass.
If axion stars collide with a white dwarf, the latter
is heated up and a detectable amount of thermal
radiation can be emitted during the collision \cite{Iw99a}.
Even stronger effects can be expected if the axion star collides
with a neutron star. If the axion star dissipates its whole
energy in a single outburst, this could be a possible mechanism
for the observed gamma ray bursts \cite{Iw99b}. For example,
a collision with a neutron star with a strong magnetic field
of about $10^{9}$ Tesla can produce gamma rays up to
$10^{21}$ eV. More details of such collisions are given in
\cite{Iw99c}. A jet of baryons and leptons
with Lorentz factors larger than 100 are generated. If the collision
does not lead to the immediate destruction of the axion star
and consists actually of several repeated collisions between
both stars, this could be the origin of time-dependent complex
properties of gamma ray bursts. A prediction for this model
is the emission of radio waves with the frequency given by the
axion mass as a byproduct to the gamma ray bursts.
In Ref.~\cite{FGU00,FGKS02}, gamma ray bursts are explained by
the relativistic detonation of electro-dilaton stars.

%*************************************
\section{Boson-fermion stars \label{BFS}}
 
In this part, we discuss the possibility of a BS located inside a fermion
star
or vice versa, depending which is the dominating component.
As an example of a fermion star, a neutron star is normally considered
as indicated by the equation of state.
So far, there are no field quantised complex scalars involved.
Since this topic was already discussed in earlier reviews we shall be
rather brief here, concentrating on new publications, and show which
type of scalar fields are applied within the combined star.

\subsection{Fermion soliton star with real scalar field\label{leepang}}
 
A combination of a BS with a fermion star
has first been  examined in the context of soliton stars \cite{LP87}.
There, a semi-classical BS with the potential
$U(\Phi)=m^2 \Phi^2 (1-\Phi/\Phi_0)^2/2$ for a real scalar
secures the existence of non-topological solitons, even in the
absence of gravity. Hence, only a conserved fermionic particle number
$N_f$ exists; that of real bosons is not well defined.
Furthermore, the real scalar interacts with the fermion field
$\psi$ through $-f \bar \psi \psi \Phi$, where $f$ is a Yukawa type
coupling constant.
For simplicity, one adopts for the fermion mass $m_f=f\Phi_0$
so that the fermion has zero effective mass inside the false
vacuum, i.e., inside the BS.
For the fermion field, a Thomas-Fermi approximation
is chosen. Remarkably, $N_f$ depends as
$M_{\rm Pl}^{9/2}/m^{3/2} \Phi_0^3$ on the mass, i.e., fractional as for
the CBS (cf.~Section \ref{critical}),
whereas the total mass $M_{\rm Pl}^{4}/m \Phi_0^2$ is similar
to that of the soliton star.
Fermion soliton BHs are mentioned in Section \ref{SBH}.
 
The observational properties of zero-mass fermion soliton stars
are investigated in \cite{Ch90a,Ch90b,Ch90c}. In Ref.~\cite{Ch90a}, only
the effects of baryons entering the star are considered if the universe
temperature drops below some critical value. For the special
choice $m_f=f\Phi_0$ of Lee and Pang, all fermions are almost
massless inside the fermion soliton star. Moreover, the binding energy
of constituent particles decrease to zero. Hence, if nuclei enter the
star, they disintegrate into protons and neutrons, and then both into
quarks. Furthermore, the fermion soliton star contents differs in the
course of universe evolution.
Above some critical temperature, protons cannot enter
(or leave), and only quark pairs are inside the star.
Below that temperature, baryons can penetrate the star's shell
and just quarks remain inside this so-called {\em quark soliton star}.
The fermion soliton stars with only quark pairs would finally
radiate away all its energy
and cease to exist; only such stars with a net number of protons
survive. These protons converts the rest energy into thermal energy so
that for a short period of time (about 10$^3$ years) at redshifts $z$
larger than 4, these stars could be X-ray emitters.
Fermion soliton stars with dominating electron part
may perturb the short wavelength character of the cosmic microwave
background radiation (CMB), causing a sharp increase in the
brightness temperature.
 
A quark soliton star at finite temperature
including the same number of (disintegrated) protons and electrons,
quark and $e^+$-$e^-$ pairs, and some so far unknown fermionic particles
with {\em ponderable} mass are calculated in \cite{Ch90b,Ch90c};
the result is labelled
{\em ponderable soliton star}. Whereas the quarks in the pure quark
soliton star are almost massless, the ponderable soliton star contains
fermions with ponderable mass. The lifetime of the ponderable soliton
star is from a few weeks to several years, they have a surface
temperature of about 10$^6$ K, and they radiate profusely
with a luminosity as high as 10$^{61}$ erg/s. Because this temperature
is close to the hydrogen recombination temperature of $\sim 10^5$ K,
these stars could be observed at redshift distances of $z\sim 10^5$.
This short radiation time  is still  longer than
the evolutionary time of the Universe at that epoch, and so,
there may be an interaction with the CMB
as well. The expected perturbations on CMB spectrum is less than 1\%
and it is largely around the short-wavelength at 1 mm or less,
in form of point radio sources; spatial inhomogeneities of arcsec
scale should be observed.
 
The situation of the fermion soliton star of Lee and Pang is extended in
two respects in Ref.~\cite{CV91a,CV91b}.
First, it is shown that the soliton star is
connected to the Lee-Wick model with scalar potential
\be
U(\Phi) = \frac{1}{2} m^2 \Phi^2 (1-\Phi/\Phi_0)^2 +
          B \Bigl[ 4-3(\Phi/\Phi_0) \Bigr] (\Phi/\Phi_0)^3\; .
\ee
{}For a constant $B^{1/4}$ of about 100 MeV as
used in hadron spectroscopy, the fermion soliton star
has a critical mass of about just three solar masses.
Additionally, the fermion soliton star at finite temperature is
investigated
and by this the formation and evolution of the stars. Above a critical
temperature around 100 MeV, the universe
is filled by the homogeneous real scalar $\Phi=\Phi_0$. Below
the critical temperature at which first-order phase transition occurs,
the real scalar adopts its  true vacuum state
$\Phi=0$; the false vacuum will only survive inside some bubbles
filled with fermions. The bubbles contract until they are stabilised
by the fermionic pressure, i.e., the fermion soliton stars have formed.
In the star's evolution, it is assumed that the star could evaporate
into hadrons either at its surface or inside via nucleated bubbles.
Then, at present, fermion soliton stars would have a mass of about
$10^{-6} M_\odot$.
 
A real scalar obeying a linear KG equation and the above mentioned
Lee-Pang
interaction are the ingredients of the model in \cite{BMH93}.
The fermionic matter has zero temperature so that one should expect no
real scalar matter at $T=0$; responsible for the occurrence of
the non-conserved scalar matter seems to be the Yukawa interaction.
In the model, instead of the exact stress-energy tensor of a Dirac
field as in \cite{OV39}, the perfect fluid approximation
for the neutron star interior is used. Actually, Ruffini and
Bonazzola \cite{RB69} proved that a large number of
degenerate fermions can be approximated by the perfect fluid approach.
Additionally, the self-interaction factor $\bar \psi \psi$
is replaced by its expectation value following \cite{RB69}.
Then, the influence of the equation of state
for an ideal Fermi gas for neutron stars is investigated, following
Chandrasekhar \cite{Ch35},
and nucleon-nucleon interactions are considered as well.
It is not surprising that the hypothetical presence of a BS
can change the cooling and the
structure of the final state of a neutron star very effectively.

%*****************
\subsection{Boson-fermion star with complex scalar field}
 
The stability of combined boson-fermion stars built from a complex scalar
is investigated in \cite{HLM90b,Je90}.
A slowly rotating boson-neutron star based on
Kaup's linear potential and Chandrasekhar's equation of state
\cite{Ch35} has been analysed in \cite{SoSi01}.
 
In Ref.~\cite{BTFY01,BTFY02}, a boson-fermion star is calculated in a
generalised JBD theory with $\varpi=0$ where the Brans-Dicke
field has a potential and, hence, a mass; for the complex scalar,
the potential $U_{\rm CSW}$ is used and solutions are
derived for $\Lambda=0.01,10$. This work can be seen as an
extension of \cite{FYBT00}; cf.~the JBD Section \ref{JBD}.

%**********************************************
\section{Perfect fluid approximation}
\label{fluid}
 
Several papers on BSs followed the old idea of Oppenheimer and Volkoff
\cite{OV39} and applied a fluid approximation for the bosonic matter,
which yields the correct order of magnitude for the total mass.
In this Section, the bosonic fluid is isotropic.
 
Nishimura and Yamaguchi \cite{NY75} constructed an equation of state
for a cold Higgs fluid
\ba
\rho & = & \frac{m^4}{4\lambda} \frac{1}{(2 u^2-1)^2} \quad (0.5 < u^2
\le 1)
  \; , \\
p    & = & \frac{m^4}{3 \lambda} \frac{1}{2 u^2-1}
 \left [ 2 \frac{E(u^2)}{K(u^2)} - \frac{2 u^2-5/4}{2 u^2-1} \right ]
 \; ,
\ea
where $m,\lambda$ are the parameters of the Higgs Lagrangian, $K,E$ the
first and second elliptic functions, respectively, and $u$ is some
parameter.
For several choices of $m^4/\lambda$, it is shown that the equation of
state $p(\rho)$ is an extrapolation of the equation of state of nuclear
matter to higher density regions; a further characteristic is that
the pressure vanishes below some value of the central density.
Substituting this equation of state into the Tolman-Oppenheimer-Volkoff
equation leads to BSs with critical masses in the order of magnitude of
a solar mass.
In a way, by taking into account a bosonic self-interaction constant,
Nishimura and Yamaguchi anticipated the result of
Colpi et al.~\cite{CSW86}
who found the critical BS masses for the Lagrangian, i.e.~anisotropic,
description in the same mass region. Furthermore, these BSs have a
finite radius, namely, the point where pressure as a function of
radius vanishes.
 
Takasugi and Yoshimura \cite{TY84} investigated a non-self-interacting
cold Bose gas with the equation of state
\be
p  =  3 m^2 n_0^2 u^{8/3}/(2\rho )\, , \qquad
\rho = m n_0 u \sqrt{1 + \frac{9}{2} u^{2/3}} \,,
\ee
where $m$ is the boson mass, $n$ the number density, $n_0$ some constant,
and $u:=n/n_0$.
We see that for small $u$, $p\propto \rho^{5/3}$ and for large
$u$, $p = \rho/3$. The critical mass is found to be 0.57 $M_{\rm Pl}^2/m$
at a radius of 5.5$/m$ which is remarkably close to the Kaup limit
0.633 $M_{\rm Pl}^2/m$ of the Lagrangian description.
 
Three different equations of state are investigated in
Ref.~\cite{TLPH88}. For massless bosons, the equation of state of photons
is used $p=\rho/3$, leading to photon stars, or ``geons".
The radius is determined where the surface pressure equals
the one of the cosmological background radiation. The critical mass is
about $10^{24}$ $M_\odot$ and the radius is larger than the
radius of the Universe,
so that it seems that photon stars cannot be found. Alternatively,
a giant MIT bag filled with gluon matter is considered, hence,
the calculation stops where the pressure is equal to the bag constant
$B$ which is chosen to be 57 MeV$/$fm$^3$. The equation of state
can be written as $p=(\rho-4B)/3$. Then, the critical mass of the gluon
star is 2 $M_\odot$. The third example describes a non-relativistic
boson gas of massive particles where Bose-Einstein condensation
is neglected, with an
equation of state $p=2(\rho-m n)/3$. Because the pressure never vanishes
in this model, the surface of these BSs is defined by $p=B$. The
critical mass is about $10^{-3}$ $M_\odot$ for $m=20$ GeV/$c^2$.
 
The MIT bag equation of state is derived in Ref.~\cite{Ha98}
for a real scalar field with self-interaction. Thereby, the effective
mass of the real scalar is zero, so that the corresponding BS
solutions are called {\em precarious stars}. These stars exist only
for temperatures above some value which is determined by the bag
constant;
this coincides with the fact that the real scalar has no
conserved charge, cf.~Section \ref{RSFBS}.
 
Departing from the Newtonian spherically symmetric Poisson equation,
Dehnen and Gensheimer \cite{DG98} analysed a self-gravitating
Bose gas cloud with temperature.
A similar situation was already considered by Ingrosso and Ruffini
\cite{IR88} for an isothermal case, i.e., constant temperature all over
the cloud; details can also be found in \cite{Je92}. The difference from
the Newtonian BSs of Section \ref{NewtBS} is that, here, an equation of
state
is derived that does not include a scalar field, but considers the
Bose-Einstein statistics in its non-relativistic limit.
It is found that the Bose gas obeys a polytropic relation
$p=a \rho^3$ with some constant $a$. The Poisson equation reduces then
to the Lane-Emden equation for polytropic index 3. In general,
it is found that the temperature
decreases with increasing distance from the centre, hence,
Bose-Einstein condensation occurs in the outer regions of the
star. This result is in contrast to the isothermal case where
the condensed phase is in the centre (additionally the outer regions
are Boltzmannian and spatially unlimited) \cite{IR88}.
Furthermore, these Bose gas clouds have a finite radius where
the energy density, the pressure, and the temperature vanish.

%***************************************
\section{Soliton black hole \label{SBH}}
 
These configurations,  introduced in \cite{FLP87b}, describe a
non-topological soliton star sitting completely inside the Schwarzschild
horizon which is possible due to the finiteness of such a star,
cf.~the radius definition in Section \ref{radii}.
That means that the radius of the star is smaller than its Schwarzschild
value.
The mass $M$ in units of $M_{\rm Pl}^4/(m\Phi_0^2)$ lies
between the values 0.1256 and $16/(27\pi)=0.1886$. For this upper bound,
the particle number $N$ vanishes.
It is not surprising that a soliton BH is not stable, it decays
very slowly by sending scalar matter towards the horizon \cite{L87b}.
Outside the horizon, the Schwarzschild solution holds; cf.~the
more general BS result in \cite{PS97}.
In \cite{PS92}, the physical properties of a soliton BH at finite
temperature was investigated following the procedure for a
non-topological soliton star in \cite{SP89}, cf.~Section \ref{hot}.
Fermion soliton BHs were constructed in \cite{LP87} for a real scalar
field;
details on fermion soliton stars in Section \ref{leepang}.

%***************************************
\section{Outlook}
 
BSs are so far hypothetical objects; what is more
it is unknown whether such localised configurations  really have the size
of a
star.
Due to the unknown nature of its  constituents,
 the dimensions of BSs could be smaller than
an atom or as large as a galactic halo. Thus, all these
different scenarios have to be considered and further
observational consequences drawn.

Within the last three decades, a lot of models have been labelled
BSs, sometimes one and the same mathematical model received
different physical designations.
All models agree that the constituents of a BS are scalars. They
differ in regards to whether the scalars are real or complex and
whether the scalars are classical or field quantised.
In this way, we can distinguish four basic types of BSs.
Furthermore, in each of these four BS types, sub-types can be
separated due to the form of the scalar self-interaction.
In this review, we tried to clarify and sum up all different boson star
notions.
 
Further analysis is needed in order to understand these highly
interesting
instances of a possible fine structure in the energy levels of general
relativistic BEC, the so-called BSs. In
view of their rich and prospective structures, are BSs
capable of explaining parts of the dark matter in the Universe?
 
We hope that this review can also be understood as a starting point
that the topic BS leaves the pure scientific area and finds its way
to popular science \cite{OS99}.
 
We think that one cannot finish a topical review on BSs any better
than by quoting from \cite{L87,FLP87a,FLP87b,LP87} and
his review \cite{LP92} the last two sentences of T.-D.~Lee:
{\em At present, there is no experimental evidence that soliton stars}
[or BSs, the authors] {\em exist. Nevertheless, it seems reasonable that
solutions of well-tested theories, such as Einstein's GR,
the Dirac equation, the KG equation, etc., should find their proper place
in nature.}

%*****************************************

\section*{Acknowledgments}
We would like to thank Carl Brans, Feodor Kusmartsev,
Diego Torres, and Andrew Whinnett
for useful discussions, literature hints, and support.
This work was partially supported by CONACyT, grants No.~3544--E9311,
No.~3898P--E9608, and by the joint German--Mexican project
DLR--Conacyt 6.B0a.6A.
One of us (F.E.S.) is partly supported by a personal fellowship.
E.W.M.~was supported by the Sistema Nacional de Investigadores (SNI).

%*****************************************

\appendix
 
\section{Convention}
 
As usual conventions can make life easier or tougher. There are
infinite possibilities to transform the BS field equations into
dimensionless quantities; two are commonplace in the literature.
In order to avoid
recalculations, we would like to propose that future authors
apply the following convention
\begin{equation}
  x = m r \; , \; \;
  \Omega = \frac{\omega}{m}\; , \; \;
 \tilde \phi_{\tt JBD}(x) = \frac{\phi_{\tt JBD}(x)}{M_{{\rm Pl}}^2} \;
,\; \;
  \sigma (x) = \sqrt{4\pi} \, \frac{P(x)}{M_{{\rm Pl}}} \; ,\; \;
  \Lambda = \frac{\lambda}{4\pi} \, \frac{M_{{\rm  Pl}}^2}{m^2} \,.
\label{dimensionless}
\end{equation}

%=======================================================================

\nonfrenchspacing
\end{document}